\documentclass[pre,aps,twocolumn,showpacs,floatfix]{revtex4-1}

\usepackage[dvips]{graphicx}
\usepackage{amsmath}
\usepackage{amssymb}
\usepackage[nice]{nicefrac}
\usepackage{color}

\begin{document}
\bibliographystyle{prsty}
\title{Stationary states in single-well potentials under symmetric L\'evy noises}

\author{Bart{\l}omiej Dybiec}
\email{bartek@th.if.uj.edu.pl}
\affiliation{M. Smoluchowski Institute of Physics, and Mark Kac Center for Complex Systems Research, Jagellonian University, ul. Reymonta 4, 30--059 Krak\'ow, Poland}

\author{Igor M. Sokolov}
\email{igor.sokolov@physik.hu-berlin.de}
\affiliation{Institut f\"ur Physik, Humboldt-Universit\"at zu Berlin, Newtonstrasse 15, D--12489 Berlin, Germany}

\author{Aleksei V. Chechkin}
\email{achechkin@kipt.kharkov.ua}
\affiliation{School of Chemistry, Tel Aviv University, Ramat Aviv, Tel Aviv 69978, Israel}
\affiliation{Institute for Theoretical Physics NSC KIPT, Akademicheskaya st. 1, Kharkov 61108, Ukraine}

\date{\today}
\begin{abstract}
We discuss the existence of stationary states for subharmonic
potentials  $V(x) \propto |x|^c$, $c<2$, under action of symmetric
$\alpha$-stable noises. We show analytically that the necessary
condition for the existence of the steady state is $c>2-\alpha$.
These states are characterized by heavy-tailed probability density
functions which decay as $P(x) \propto x^{-(c+\alpha -1)}$ for $|x|
\to \infty$, i.e. stationary states posses a heavier tail than the corresponding
$\alpha$-stable law. Monte Carlo simulations confirm the existence
of such stationary states and the form of the tails of corresponding
probability densities.
\end{abstract}

\pacs{
    05.40.Fb, 
    05.10.Gg, 
    02.50.-r, 
    02.50.Ey, 
    }
\maketitle

\section{Introduction\label{sec:introduction}}

The dynamics of many physical systems can be reduced to a standard
model of a ``particle'' (relevant coordinate $x$) under action of a
deterministic force $f(x)$ and the noise force
including the action of all neglected degrees of freedom. The
mathematical tool for such a description is given by a Langevin
equation, which in the overdamped limit typically takes the form
\begin{equation}
\dot{x}(t)=f(x)+\zeta(t).
\label{eq:langevin0}
\end{equation}
In Eq.~(\ref{eq:langevin0}) $f(x)$ stands for the deterministic
force, while $\zeta(t)$ is the stochastic force --- noise ---
describing interactions of a test particle with its complex
surrounding. Commonly, it is assumed that the stochastic force is of
the white type, for example representing the particle's collisions
with molecules of the bath. Following Generalized Central Limit
Theorem, a large number of independent collisions will lead to white
noise of the stable type, i.e. either to Gaussian noise or to more
general L\'evy noise. The presence of fluctuations distributed
according to L\'evy laws have been observed in various situations in
physics, chemistry or biology \cite{shlesinger1995,nielsen2001},
paleoclimatology \cite{ditlevsen1999b} or economics
\cite{mantegna2000}. The situations pertinent to heavy-tailed
distributions appear in context of different models
\cite{solomon1993,solomon1994,chechkin2002b,boldyrev2003}, and are
analyzed in an increasing number of studies
\cite{jespersen1999,ditlevsen1999,kosko2001,dybiec2004,delcastillonegrete2008,dybiec2004b,dybiec2006,dybiec2006b,dybiec2007,dybiec2009,chechkin2002,chechkin2003,chechkin2004,sokolov2003,dubkov2008}.

The action of white L\'evy noise leads to increments of the
stochastic process which are distributed according to
$\alpha$-stable L\'evy type distributions. Symmetric L\'evy
distributions $p_{\alpha}(x;\sigma)$ are characterized by their
Fourier-transforms (characteristic functions of the distributions)
$\phi(k) = \int_{-\infty}^\infty e^{ikx} p_{\alpha}(x;\sigma) dx$
being \cite{feller1968,janicki1994,janicki1996}
\begin{equation}
\phi(k) = \exp\left[ -\sigma^\alpha|k|^\alpha\right].
\label{eq:charakt}
\end{equation}
The parameter $\alpha$ (where $\alpha\in(0,2]$) is the stability index
of the distribution describing, for $\alpha <2$, the asymptotic decay of its tails
\begin{equation}
p_{\alpha}(x) \propto |x|^{-(1+\alpha)}.
\label{eq:asymptotics}
\end{equation}
Finally, $\sigma$ is the scale parameter which controls the overall distribution width.
The Gaussian distribution corresponds to a special case of a L\'evy stable law with
$\alpha=2$ when $\sqrt{2} \sigma$ is interpreted as the standard deviation of the distribution.
In more general cases, for $\alpha<2$, the variance of $\alpha$-stable densities diverges. For $\alpha<1$, the mean value of $\alpha$-stable densities also does not exist.

The present work addresses properties of L\'evy flights in external
potentials. Theoretical descriptions of such systems is based on the
Langevin equation and/or Fokker-Planck equation, which is of the
fractional order \cite{podlubny1998}.  The research performed here
extends earlier studies
\cite{jespersen1999,chechkin2002,chechkin2003,chechkin2004} where
analysis of L\'evy flights in harmonic and superharmonic potentials
has been presented. The discussion conducted there did not account
for the problem of existence of stationary states in potential less
steep than parabolic, which is the main topic of the present work.
This issue is addressed here by the use of analytical arguments and
Monte Carlo simulations.

The model under discussion is presented in
Section~\ref{sec:model}. Section~\ref{sec:results} discusses obtained
results. The paper is closed with concluding remarks
(Section~\ref{sec:summary}).

\section{Model\label{sec:model}}

%
%
\begin{figure}
\begin{center}
\includegraphics[angle=0,width=8cm]{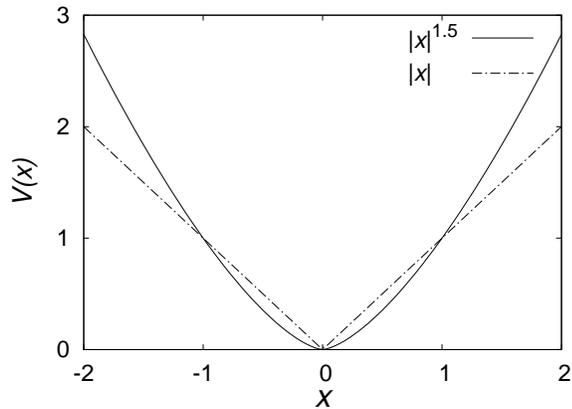}
\caption{Exemplary subharmonic potentials: $V(x)=|x|$ and $V(x)=|x|^{1.5}$ used for examination of the problem of existence of stationary states.}
\label{fig:potential}
\end{center}
\end{figure}

An overdamped Brownian-L\'evy particle moves in an external potential $V(x)$, therefore Eq.~(\ref{eq:langevin0}) takes the form
\begin{equation}
\dot{x}(t)=-V'(x)+\zeta(t),
\label{eq:langevin}
\end{equation}
where $V(x)$ represents a subharmonic single well potential $V(x)=|x|^c$ with $0<c<2$, see Fig.~\ref{fig:potential}, and $\zeta(t)$ denotes a L\'evy stable white noise process
\cite{weiss1983,caceras1999,jespersen1999,ditlevsen1999,dybiec2006}. The
position of a random walker can be calculated by means of the stochastic integration of Eq.~(\ref{eq:langevin}) \cite{janicki1994,janicki1994b}
\begin{eqnarray}
x(t)& = & x(0)-\int_0^t V'(x(s))ds + \int_0^t\zeta(s)ds \nonumber \\
& = & x(0)-\int_0^t V'(x(s))ds + L_\alpha(t).
\end{eqnarray}
The integral $\int_0^t\zeta(s)ds\equiv L_\alpha(t)$
defines an $\alpha$-stable L\'evy process $L_\alpha(t)$
\cite{weiss1983,samorodnitsky1994,jespersen1999,ditlevsen1999,dybiec2006} which is driven
by a L\'evy stable noise $\zeta(t)$. Increments $\Delta L_\alpha(\Delta t) = L_\alpha(t + \Delta t)- L_\alpha(t)$ of the $\alpha$-stable L\'evy process  are distributed
according to the $\alpha$-stable density with the stability index $\alpha$.
If the time step is set to $\Delta t$, the appropriate $\alpha$-stable distribution is
$p_{\alpha}(\Delta L_\alpha;\sigma(\Delta t)^{1/\alpha})$
\cite{janicki1994,janicki1996,janicki2001,nolan2002}.

The equation (\ref{eq:langevin}) is associated
with the following fractional Fokker-Planck equation (FFPE)
\cite{metzler1999,yanovsky2000,schertzer2001,paola2003,brockmann2002}
\begin{eqnarray}
\label{eq:ffpe}
\frac{\partial P(x,t)}{\partial t} & = & \left[ \frac{\partial}{\partial x} V'(x,t) + \sigma^\alpha \frac{\partial^\alpha}{\partial |x|^\alpha}\right]P(x,t)
\end{eqnarray}
where the fractional (Riesz-Weyl) derivative is defined by the
Fourier transform
\cite{jespersen1999,chechkin2002,metzler2004,chechkin2003}
$\mathcal{F}\left[ \frac{\partial^\alpha}{\partial|x|^\alpha}f(x)
\right]=-|k|^\alpha \mathcal{F}\left[ {f}(x) \right].$ The (space)
fractional derivative in Eq.~(\ref{eq:ffpe}) describes long jumps
which are distributed according to the $\alpha$-stable density, see
Eq.~(\ref{eq:charakt}) and
Refs.~\cite{shlesinger1995,yanovsky2000,paola2003,dubkov2005b}. The
Langevin equation~(\ref{eq:langevin}) provides a stochastic
representation of the fractional Fokker-Planck
equation~(\ref{eq:ffpe}). In other words, it describes an evolution
of a single realization of the stochastic process $\{ x(t) \}$. From
ensemble of trajectories it is possible to analyze further
properties of the system, in particular a probability density
$P(x,t)$ of finding a random walker at time $t$ in the neighborhood
of $x$, see Eq.~(\ref{eq:ffpe}). More details on numerical scheme of
integration of stochastic differential equations with respect to
$\alpha$-stable noises can be found in
\cite{janicki1994,janicki1996,janicki2001,dybiec2004b,dybiec2006}.

In the following sections properties of stationary probability distributions for
single-well systems perturbed by symmetric L\'evy noises are discussed. In the limiting cases of parabolic and quartic potentials the performed
simulations corroborate earlier theoretical findings
\cite{jespersen1999,chechkin2002,chechkin2003,chechkin2004,chechkin2006}.

\section{Results \label{sec:results}}

%
%
\begin{figure}
\begin{center}
\begin{tabular}{p{0.45\columnwidth}p{0.45\columnwidth}}
\includegraphics[angle=0,width=0.45\columnwidth]{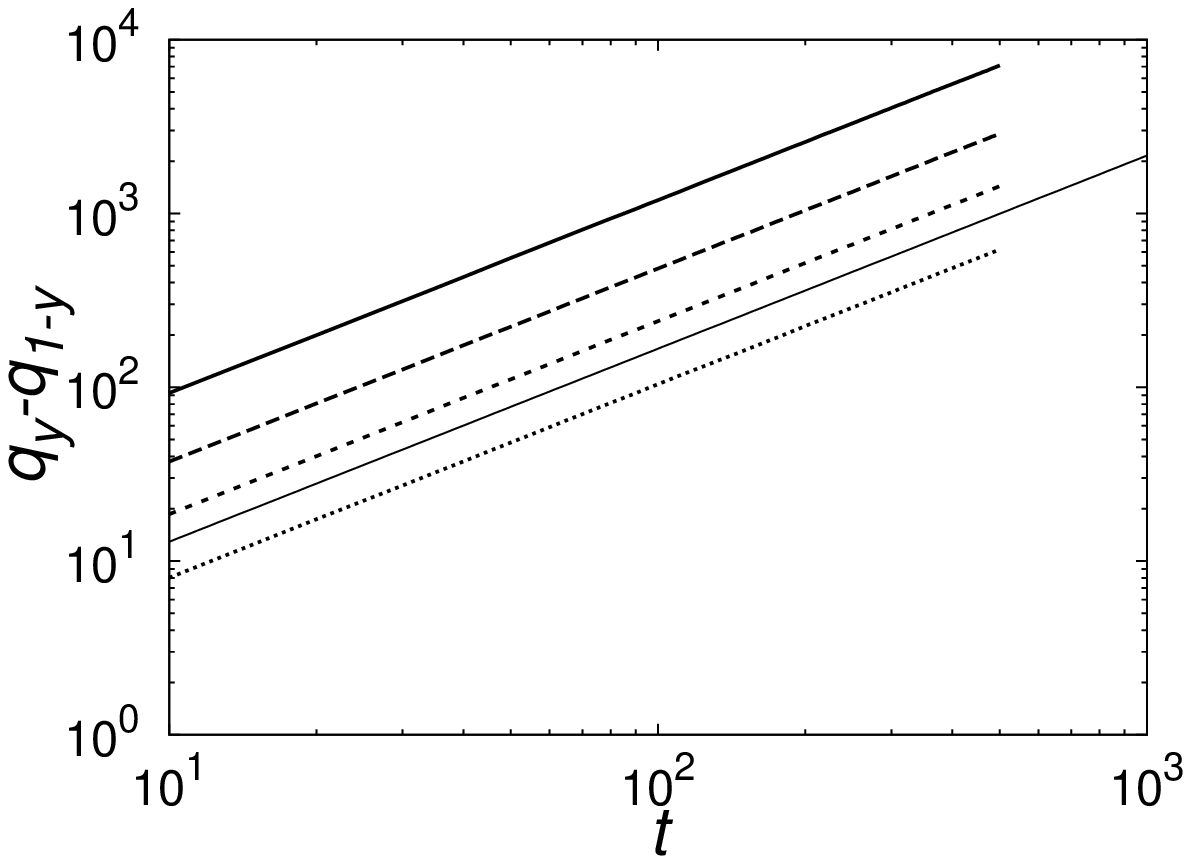} & \includegraphics[angle=0,width=0.45\columnwidth]{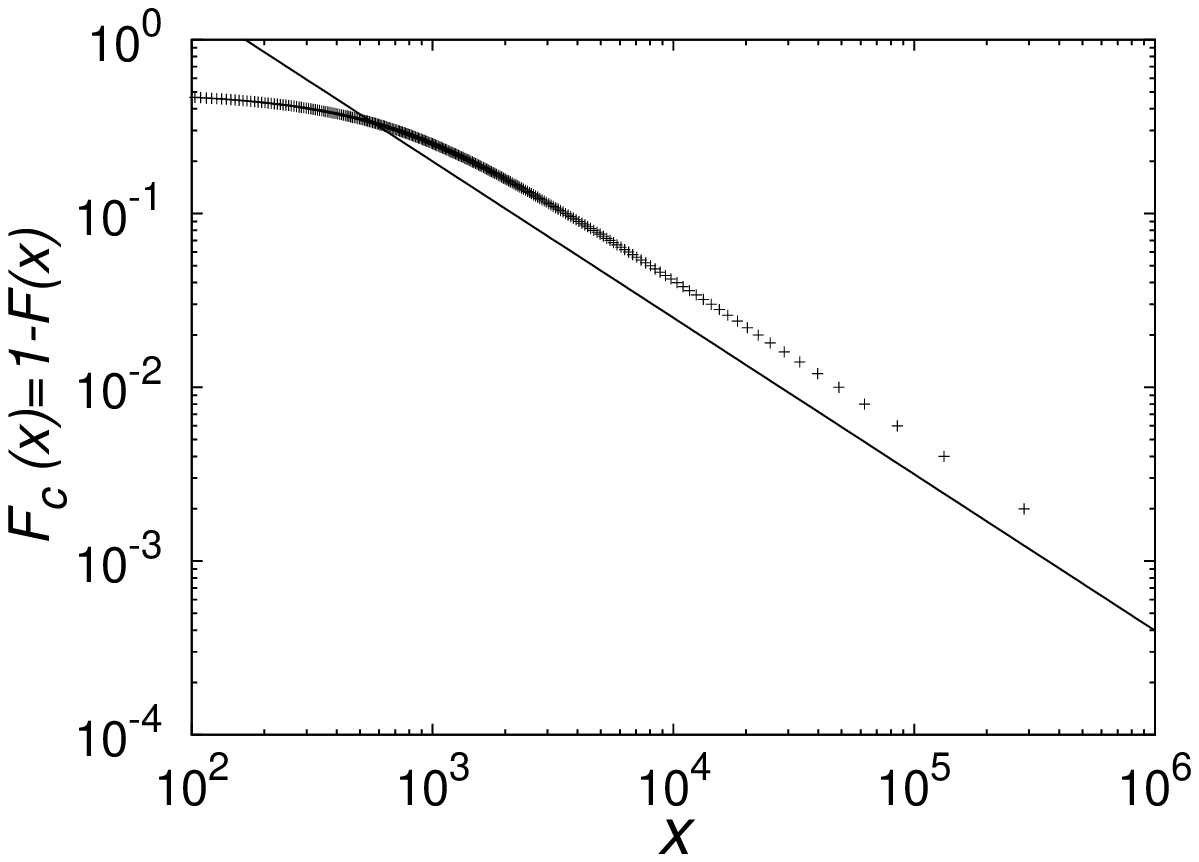}\\
\includegraphics[angle=0,width=0.45\columnwidth]{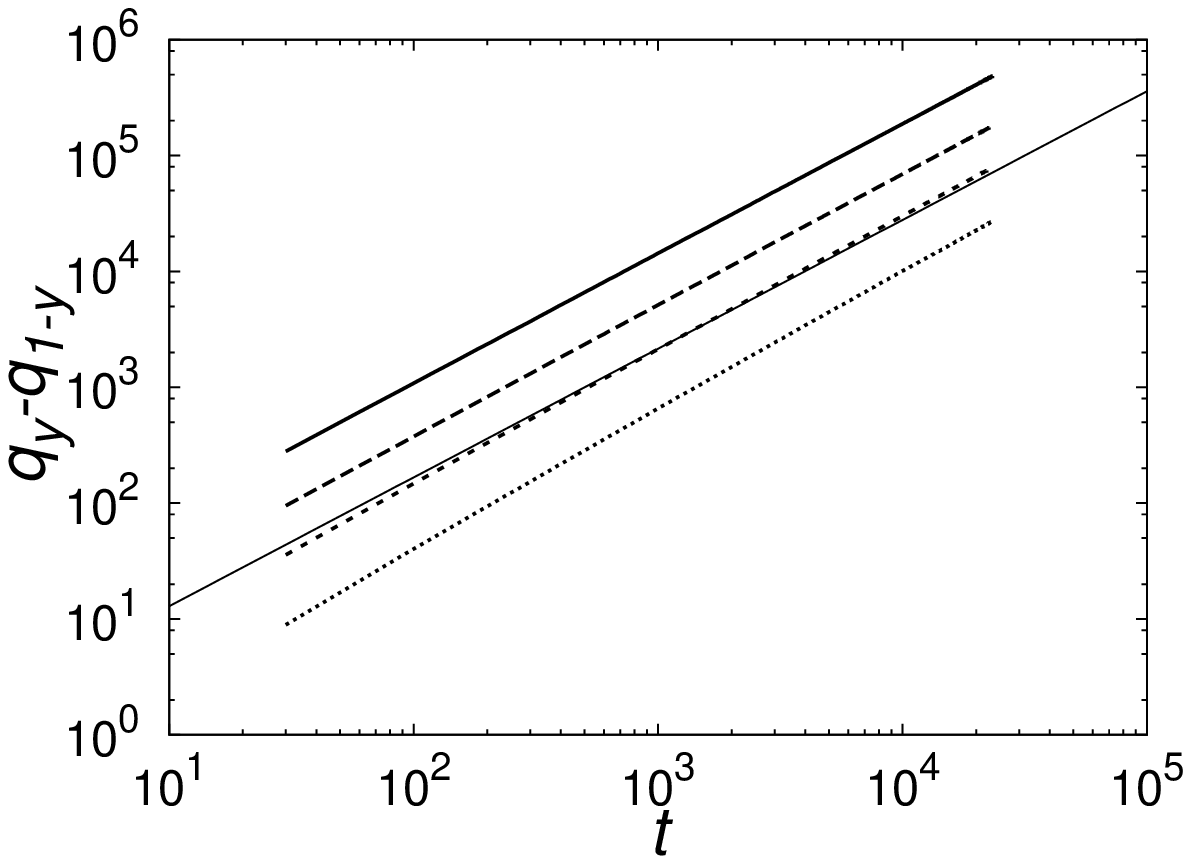} & \includegraphics[angle=0,width=0.45\columnwidth]{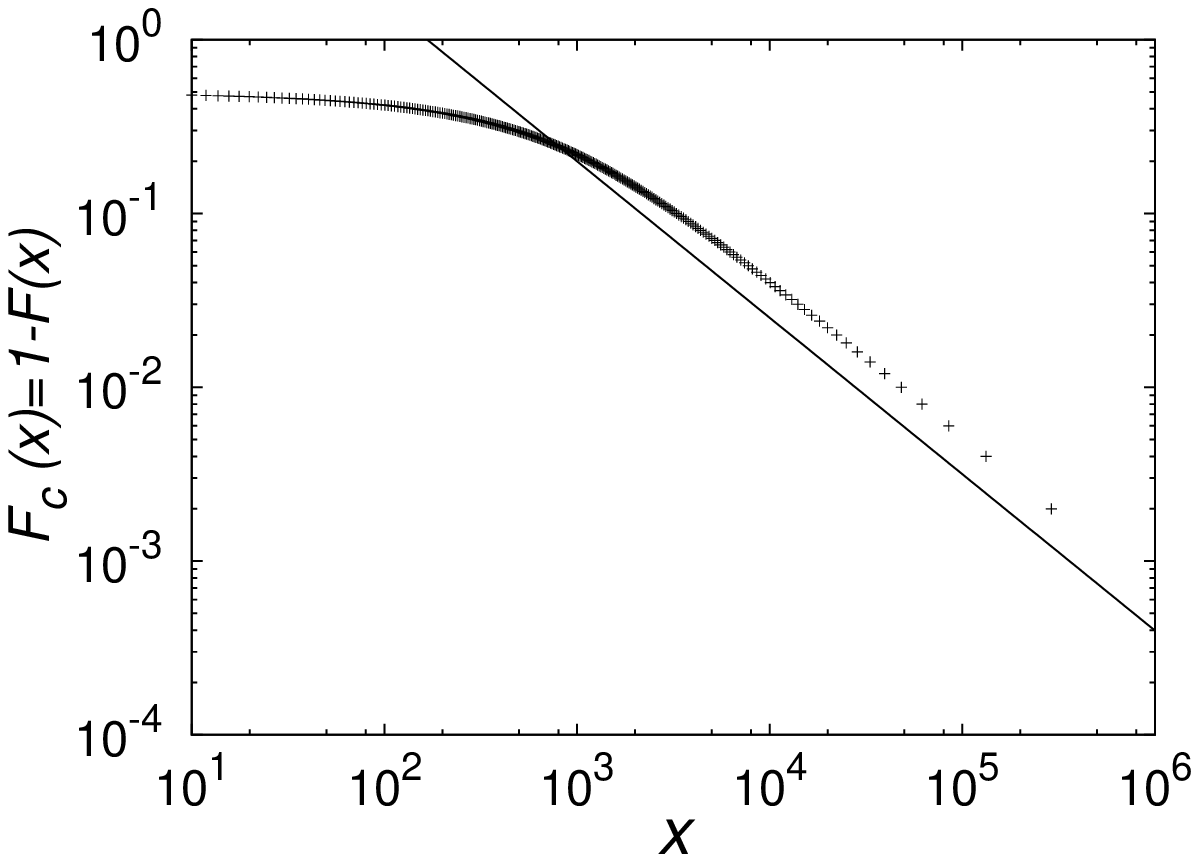}\\
\includegraphics[angle=0,width=0.45\columnwidth]{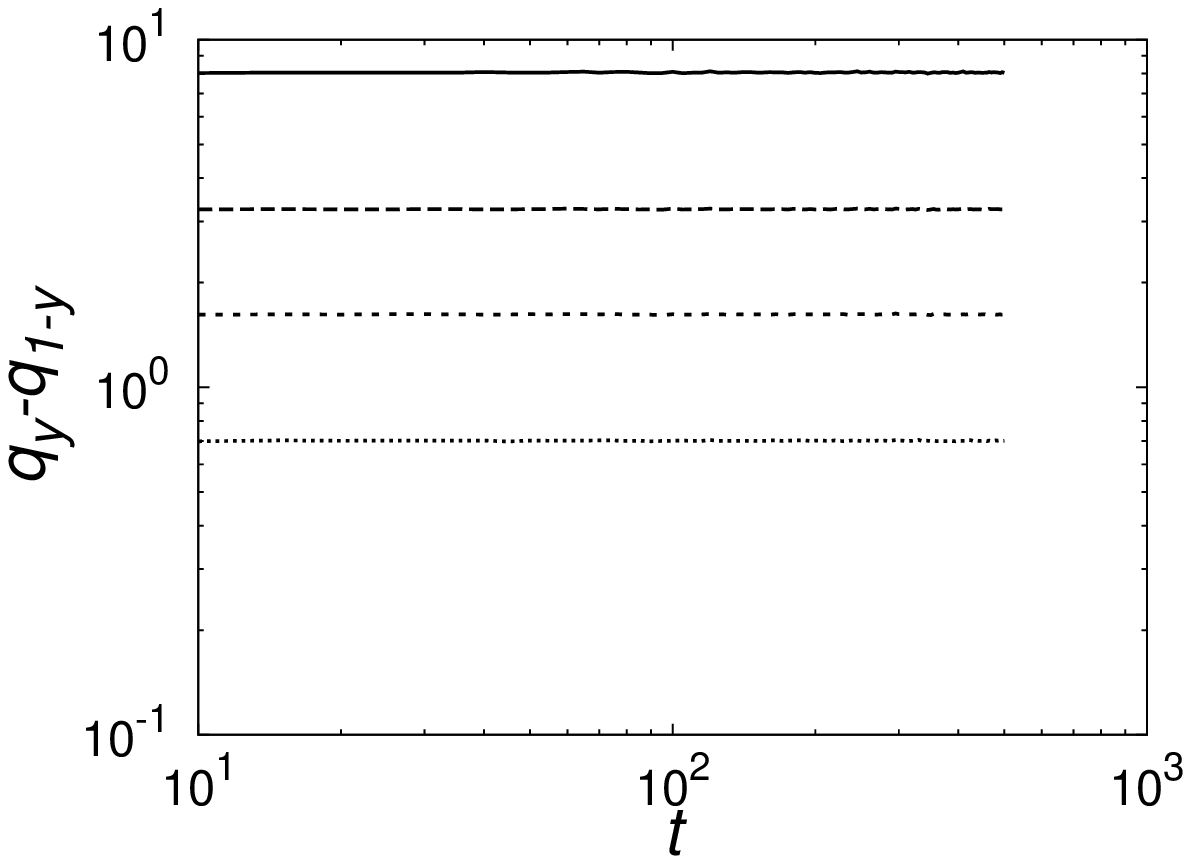} & \includegraphics[angle=0,width=0.45\columnwidth]{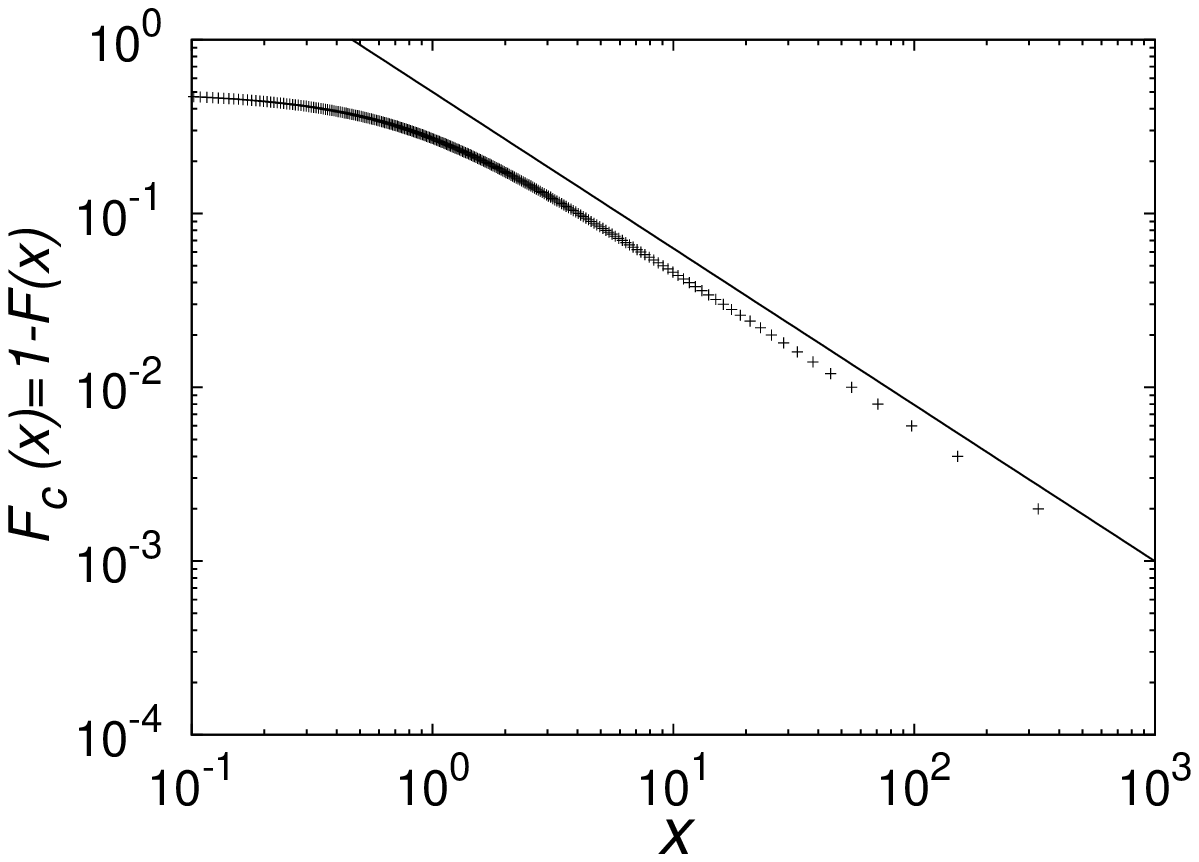}\\
\end{tabular}
\caption{Left column presents interquantile distance
($q_{0.9}-q_{0.1}$, $q_{0.8}-q_{0.2}$, $q_{0.7}-q_{0.3}$ and
$q_{0.6}-q_{0.4}$, from top to bottom) as a function of time $t$.
Solid lines present $t^{1/\alpha}$ scaling of the interquantile
distance, which is observed in the force free case. Right column
demonstrates complementary cumulative distributions at the end of
simulation. Solid lines present $x^{-\alpha}$ decay. Various rows
correspond to different potentials: $V(x)=0$ (top panel), $V(x)=|x|$
(middle panel) and $V(x)=x^2/2$ (bottom panel). The stability index
$\alpha$ and the scale parameter $\sigma$ are set to $\alpha=0.9$
and $\sigma=1$ respectively. } \label{fig:potentials}
\end{center}
\end{figure}

Only in the limited number of special cases it is possible to find
stationary probability densities $P(x)$ analytically by solving
Eq.~(\ref{eq:ffpe}).  Therefore, the construction of stationary
densities rely on numerical methods.  In such a case, there are two
frameworks possible: one option is to discretize Eq.~(\ref{eq:ffpe}), see
\cite{press1992,chechkin2002,chechkin2004,abdelrehim,meerschaert2004,meerschaert2006b},
alternatively one might use a Monte-Carlo method based on the
simulation of the Langevin equation~(\ref{eq:langevin}), see
\cite{janicki1994,janicki1994b,janicki1996,janicki2001,dybiec2007d}.  Here, we
rely on Monte Carlo simulations. Such a choice is based on statistical
properties of searched densities.  As we proceed to show, for
subharmonic potentials, if stationary states exist, the tail of the
corresponding probability density function $P(x)$ is ``fatter'' than
the tail of the corresponding $\alpha$-stable density. In such a case
confining the numerical solution of equation (\ref{eq:ffpe}) to a
finite interval may introduce uncontrollable errors. Therefore, in our
simulations, we rely solely on stochastic representation of the
fractional Fokker-Planck equation (\ref{eq:ffpe}) which is provided by
Eq.~(\ref{eq:langevin}).

%
%
\begin{figure}
\begin{center}
\begin{tabular}{p{0.45\columnwidth}p{0.45\columnwidth}}
\includegraphics[angle=0,width=0.45\columnwidth]{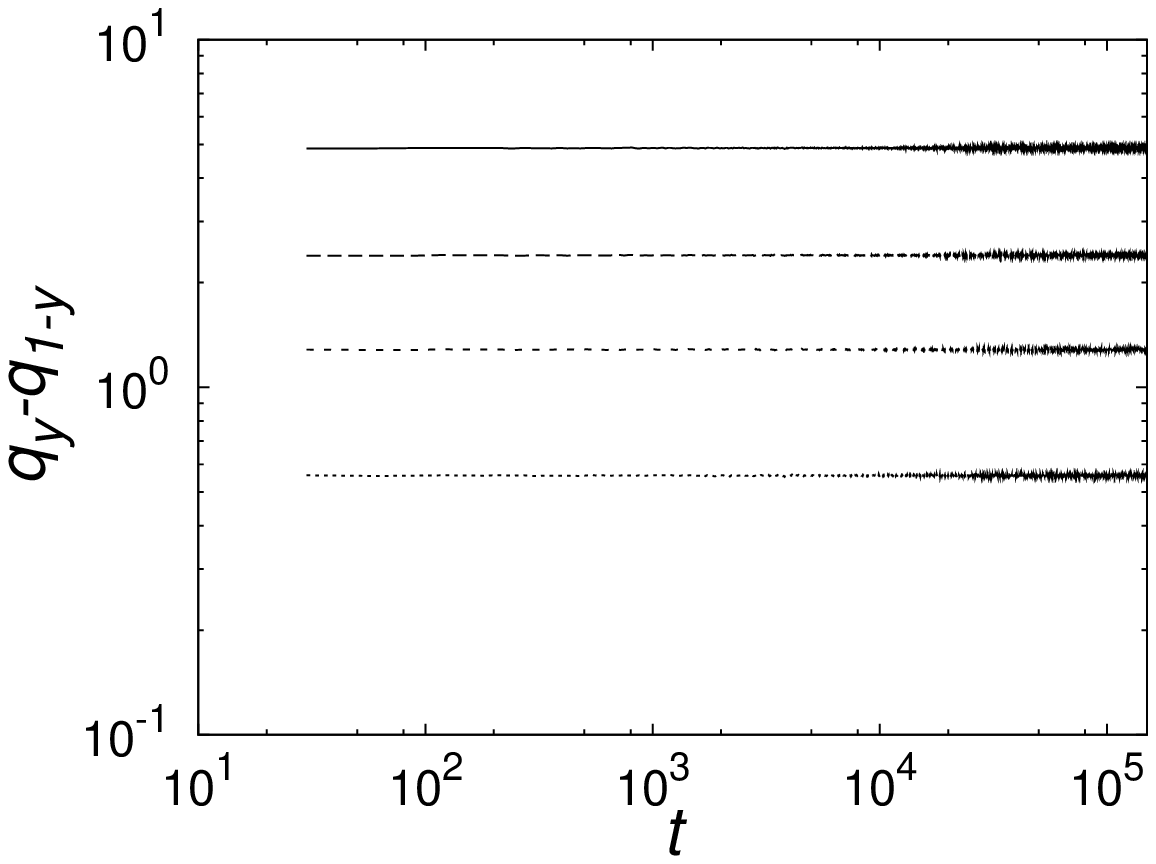} & \includegraphics[angle=0,width=0.45\columnwidth]{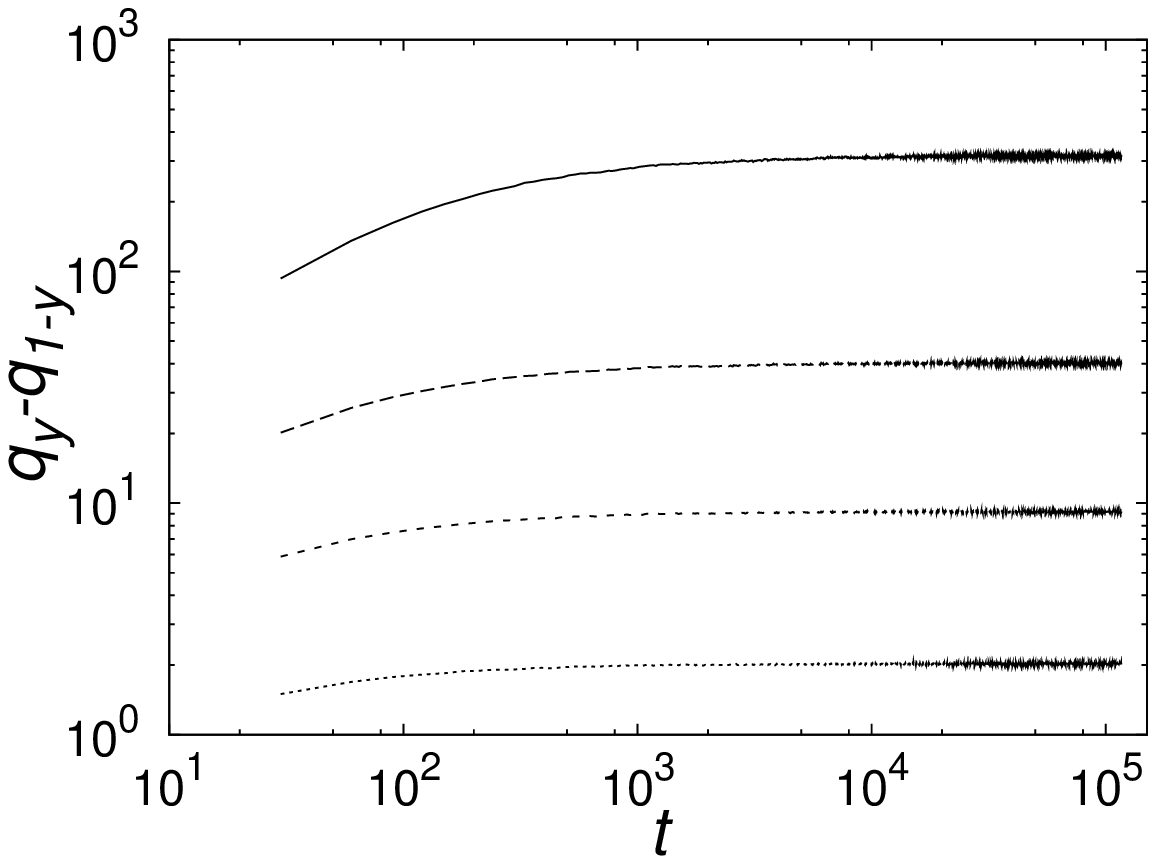}\\
\includegraphics[angle=0,width=0.45\columnwidth]{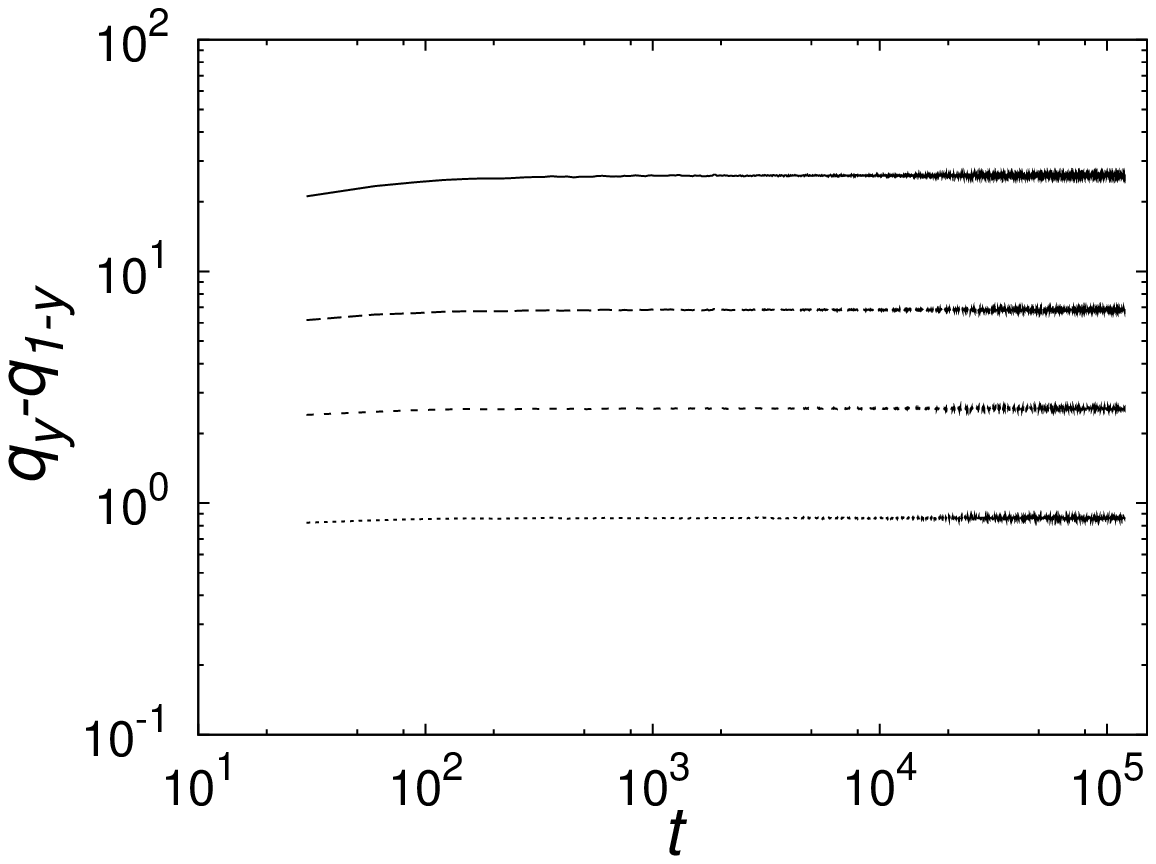} & \includegraphics[angle=0,width=0.45\columnwidth]{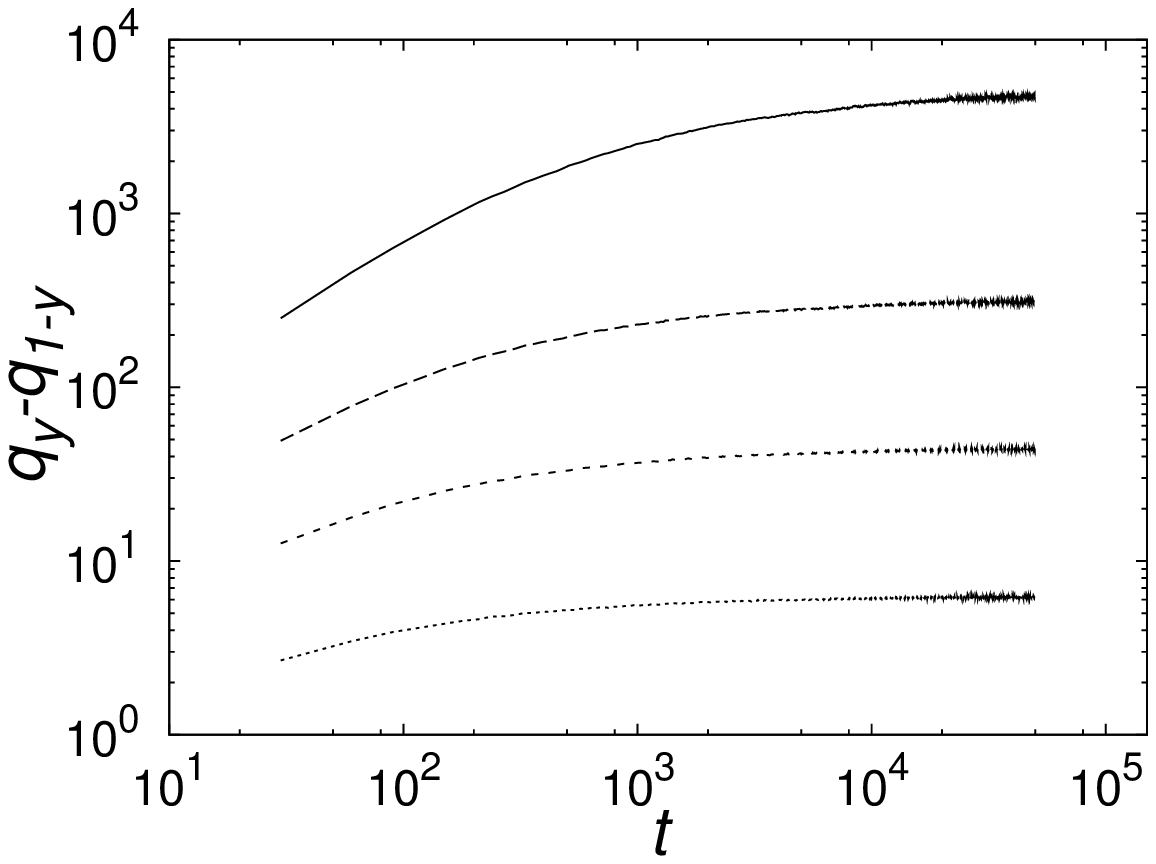}\\
\end{tabular}
\caption{Interquantile distance ($q_{0.9}-q_{0.1}$, $q_{0.8}-q_{0.2}$, $q_{0.7}-q_{0.3}$ and $q_{0.6}-q_{0.4}$, from top to bottom) as a function of time $t$ for $V(x)=|x|^{1.5}$ with different values of the stability index $\alpha=\{1.5,1.1\}$ (left column, from top to bottom) and $\alpha=\{0.9,0.8\}$ (right column, from top to bottom). The scale parameter $\sigma$ is set to $\sigma=1$.}
\label{fig:x15intqnt}
\end{center}
\end{figure}

In the $\alpha=2$ case (Gaussian noise)
the stationary solution for Eq.~(\ref{eq:ffpe}), which in this case reduces to a ``normal'' Fokker-Planck equation,
has the Boltz\-mann-Gibbs form
\begin{equation}
P(x) \propto \exp\left[-\frac{V(x)}{\sigma^2} \right]
\label{eq:gibbspdf}
\end{equation}
and exists for any power-law potential $V(x)$ such that $\lim_{|x|\to \infty}V(x)=+\infty$.
The situation drastically changes for L\'evy noises with $\alpha<2$.
Stationary states always exist in
potentials steeper than the parabolic one \cite{chechkin2002,chechkin2003,chechkin2004,chechkin2006}. However,
as it will be shown this is no more the case for subharmonic potentials.


\subsection{Parabolic and quartic potentials\label{sec:parabolic}}
There are two special cases of power-law potentials for which the stationary solution can be given in a closed form:
the case of a parabolic potential (for any $0<\alpha \leqslant 2$) and the case of a quartic potential for $\alpha=1$ and $\alpha=2$.
These cases will serve as a benchmark for our further considerations.

In order to find stationary solution of the  fractional Fokker-Planck equation (\ref{eq:ffpe}) we rewrite it in the Fourier space
\begin{equation}
\frac{\partial \hat{P}(k,t)}{\partial t} = \hat{U}(k)
\hat{P}(k,t) - \sigma^\alpha |k|^\alpha \hat{P}(k,t).
\label{eq:ftstationary}
\end{equation}
In Eq.~(\ref{eq:ftstationary}) $\hat{U}(k)$ denotes the operator
related to the Fourier representation of the potential $V(x)$ \cite{chechkin2004}.

For the parabolic potential $V(x)=x^2/2$ the expression for $\hat{U}(k)$
reads $\hat{U}(k)=-k\frac{\partial}{\partial k}$. For symmetric
$\alpha$-stable noises, the stationary probability density fulfills
\cite{chechkin2002,chechkin2003}
\begin{equation}
\frac{\partial\hat{P}(k)}{\partial k}=-\sigma^\alpha\mathrm{sign}k|k|^{\alpha-1}\hat{P}(k).
\label{eq:parabolicft}
\end{equation}
The solution of Eq.~(\ref{eq:parabolicft}) is
\begin{equation}
\hat{P}(k)=\exp\left[- \frac{\sigma^\alpha|k|^\alpha}{\alpha}\right],
\label{eq:parabolic}
\end{equation}
i.e., the stationary solution is a symmetric L\'evy distribution, see
Eq.~(\ref{eq:charakt}), characterized by a different value of the
scale parameter than the noise in Eq.~(\ref{eq:langevin}). The scale
parameter of the stationary density is $\sigma/\alpha^{1/\alpha}$. For $\alpha<2$, the
variance of the stationary solution diverges and the parabolic
potential is not sufficient to produce bounded states, i.e. states
characterized by a finite variance
\cite{jespersen1999,chechkin2002,chechkin2003,chechkin2004,dubkov2005b}.

%
%
\begin{figure}
\begin{center}
\begin{tabular}{p{0.45\columnwidth}p{0.45\columnwidth}}
\includegraphics[angle=0,width=0.45\columnwidth]{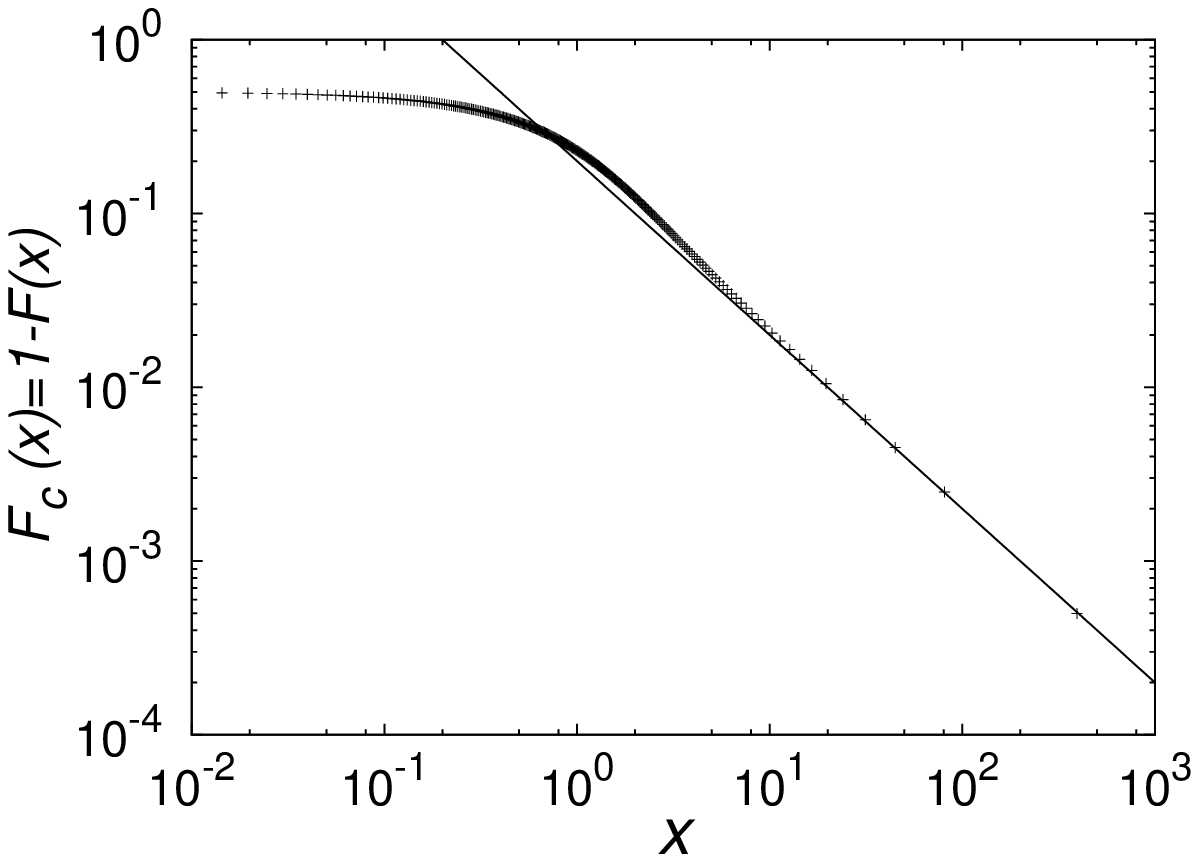} & \includegraphics[angle=0,width=0.45\columnwidth]{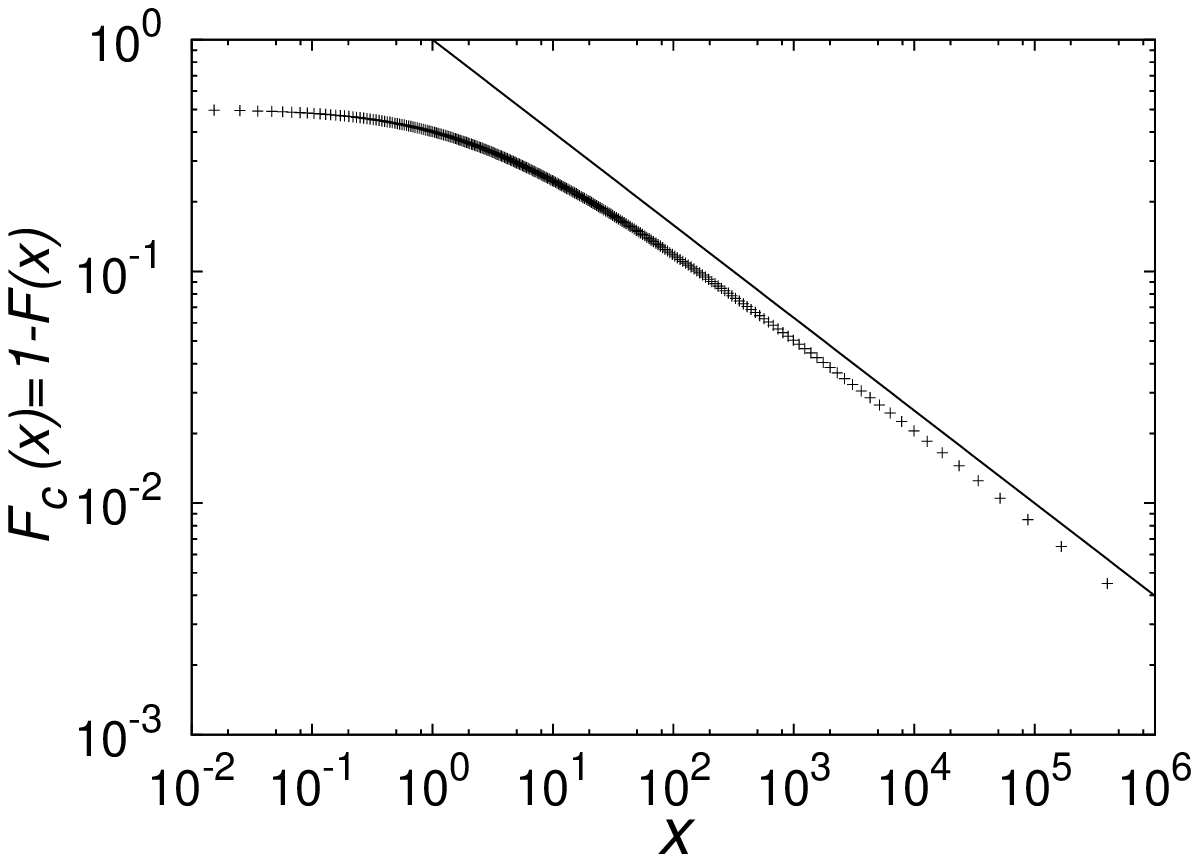} \\
\includegraphics[angle=0,width=0.45\columnwidth]{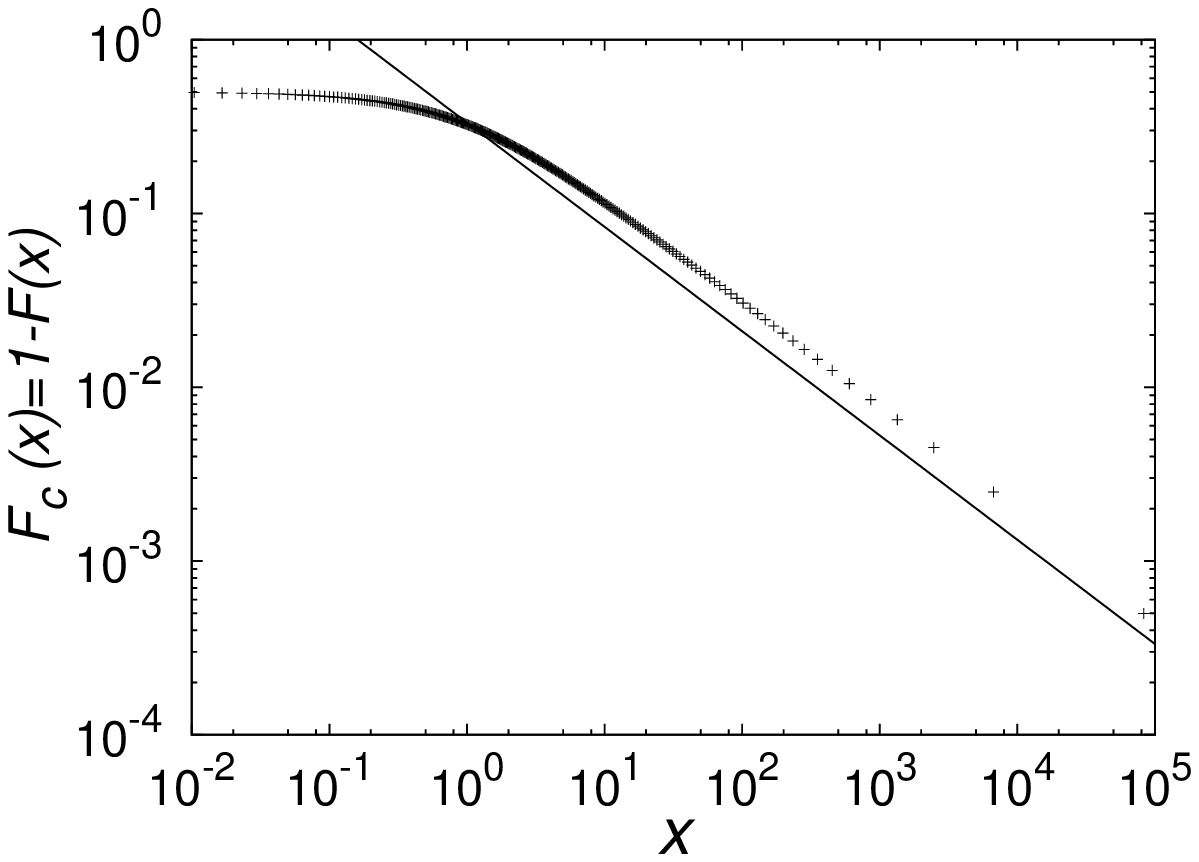} & \includegraphics[angle=0,width=0.45\columnwidth]{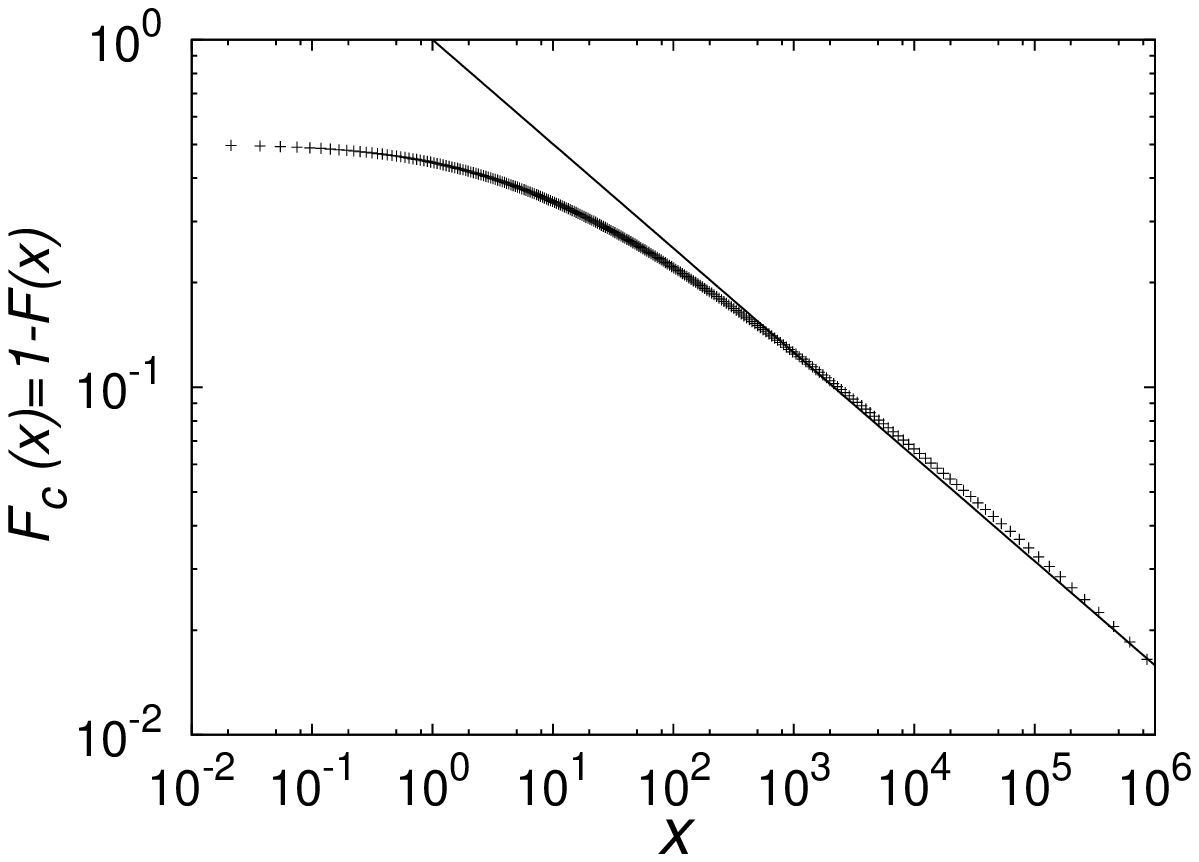} \\
\end{tabular}
\caption{Complementary cumulative distributions at the end of
simulation for $V(x)=|x|^{1.5}$ with different values of the
stability index $\alpha=\{1.5,1.1\}$ (left column, from top to
bottom) and $\alpha=\{0.9,0.8\}$ (right column, from top to bottom).
The scale parameter $\sigma$ is set to $\sigma=1$.  Solid lines
present $x^{-(\alpha-0.5)}$ decay predicted by
Eq.~(\ref{eq:hdfdecay}).} \label{fig:x15cdf}
\end{center}
\end{figure}

%
%
\begin{figure}
\begin{center}
\begin{tabular}{p{0.45\columnwidth}p{0.45\columnwidth}}
\includegraphics[angle=0,width=0.45\columnwidth]{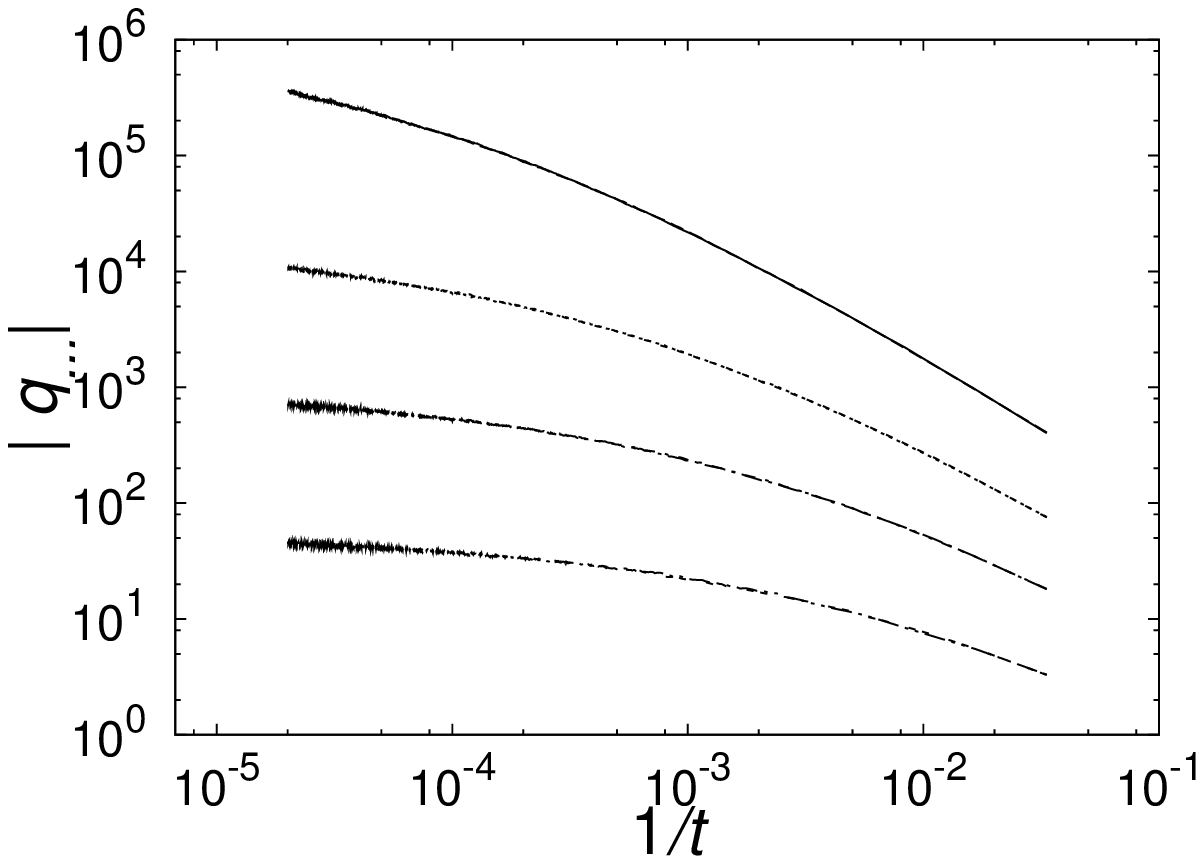} & \includegraphics[angle=0,width=0.45\columnwidth]{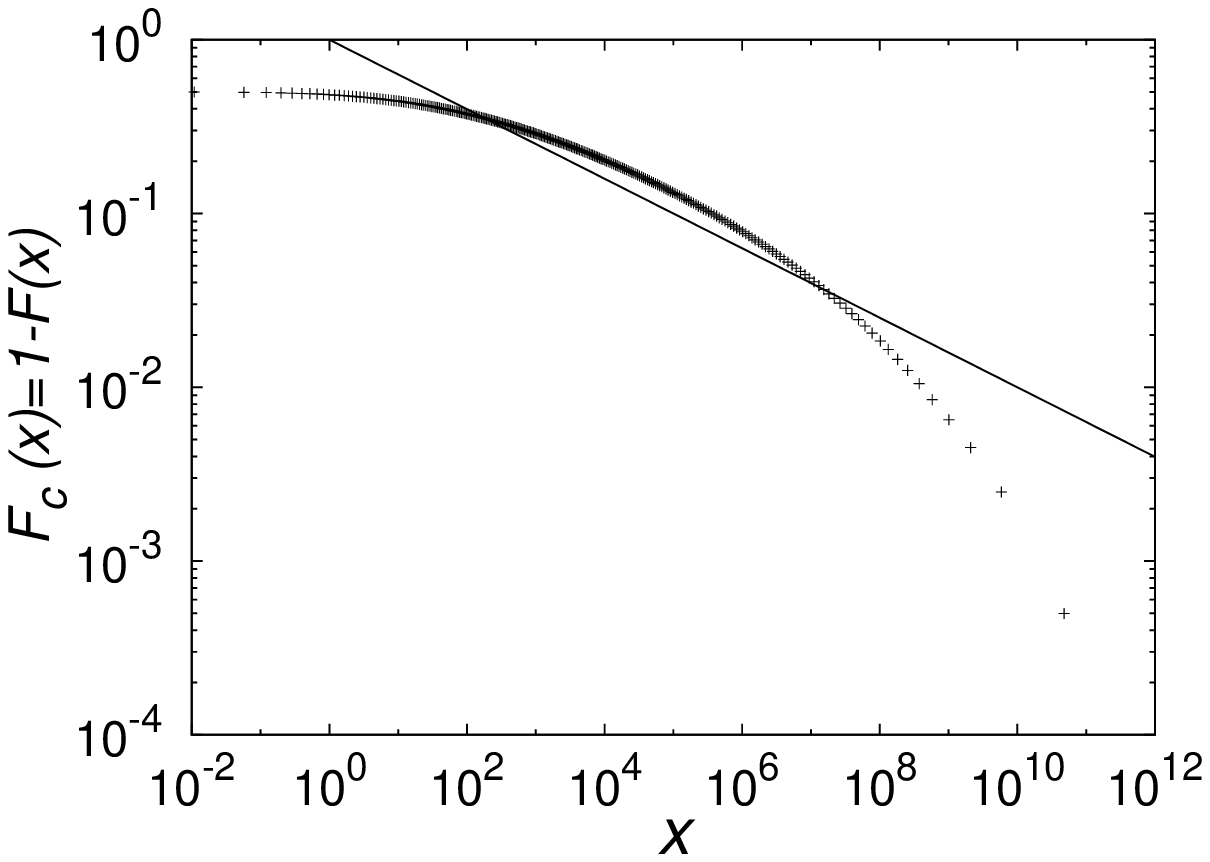}\\
\includegraphics[angle=0,width=0.45\columnwidth]{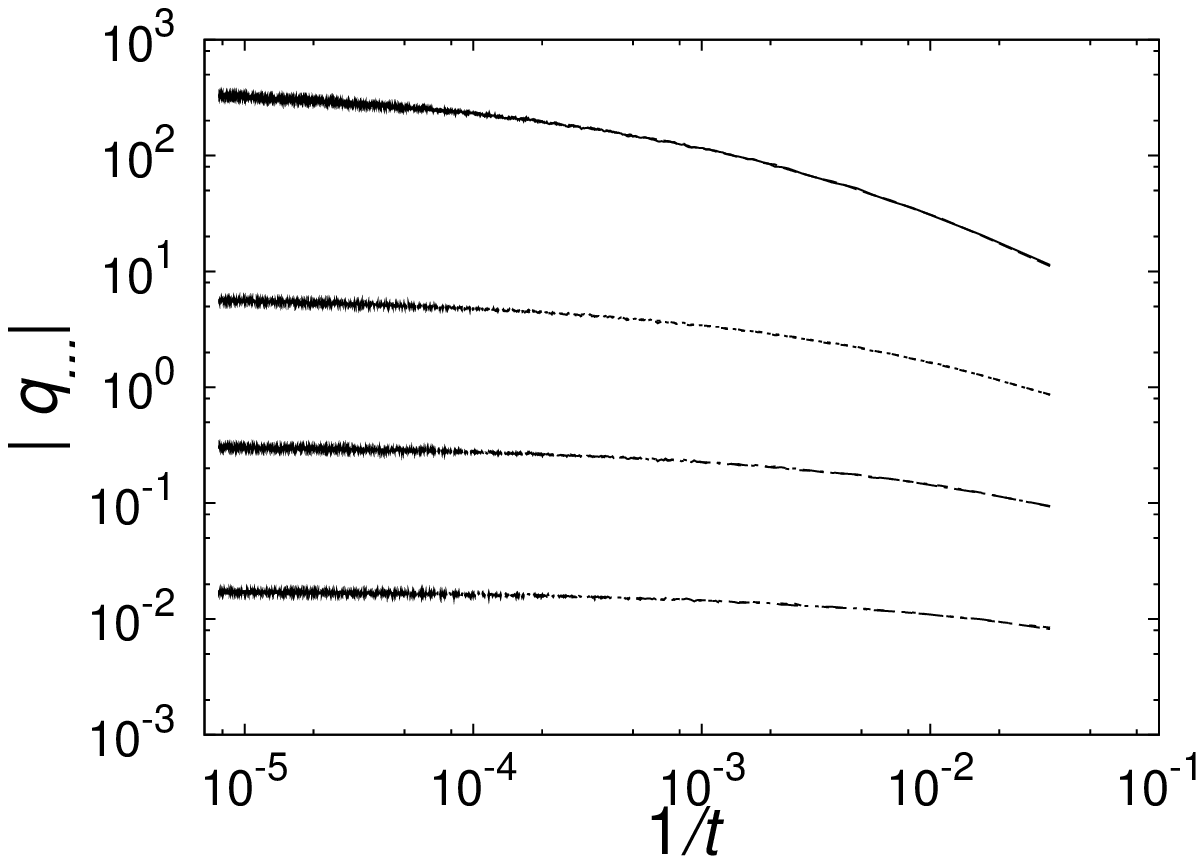} & \includegraphics[angle=0,width=0.45\columnwidth]{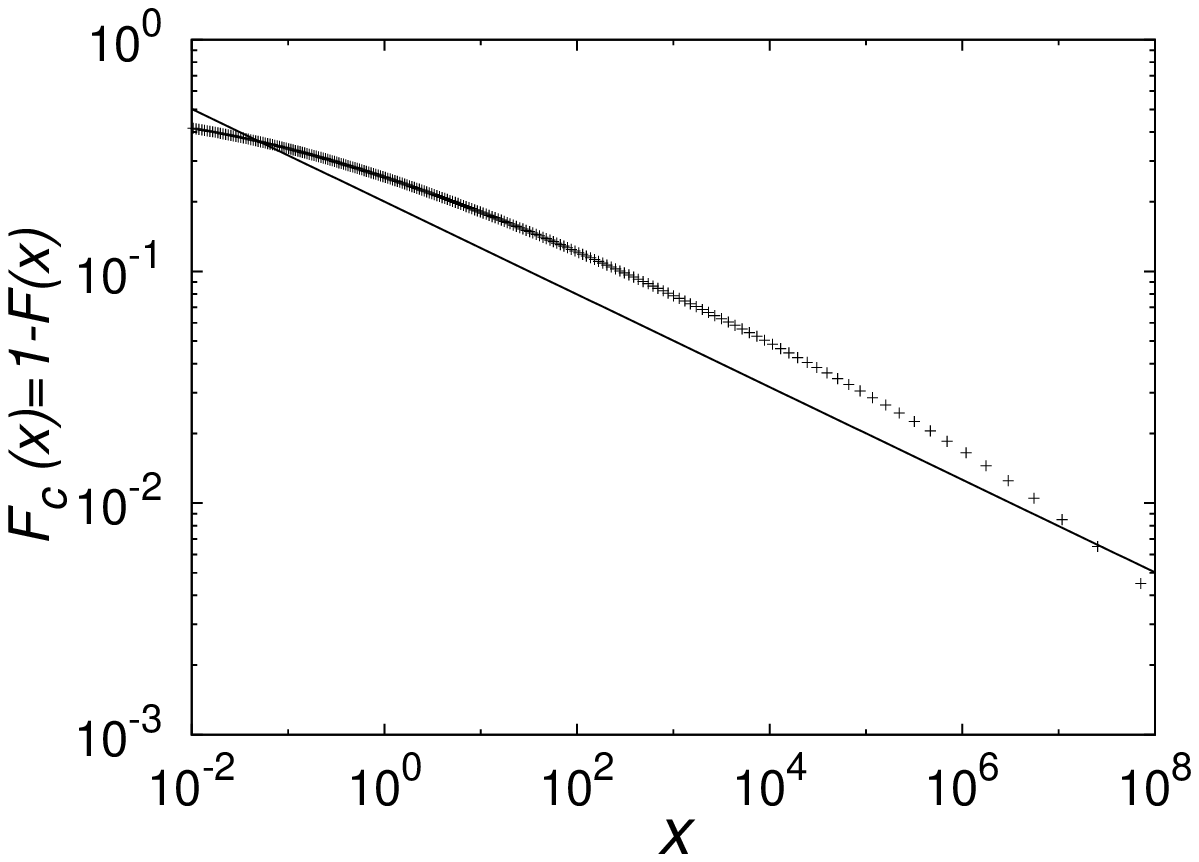}\\
\includegraphics[angle=0,width=0.45\columnwidth]{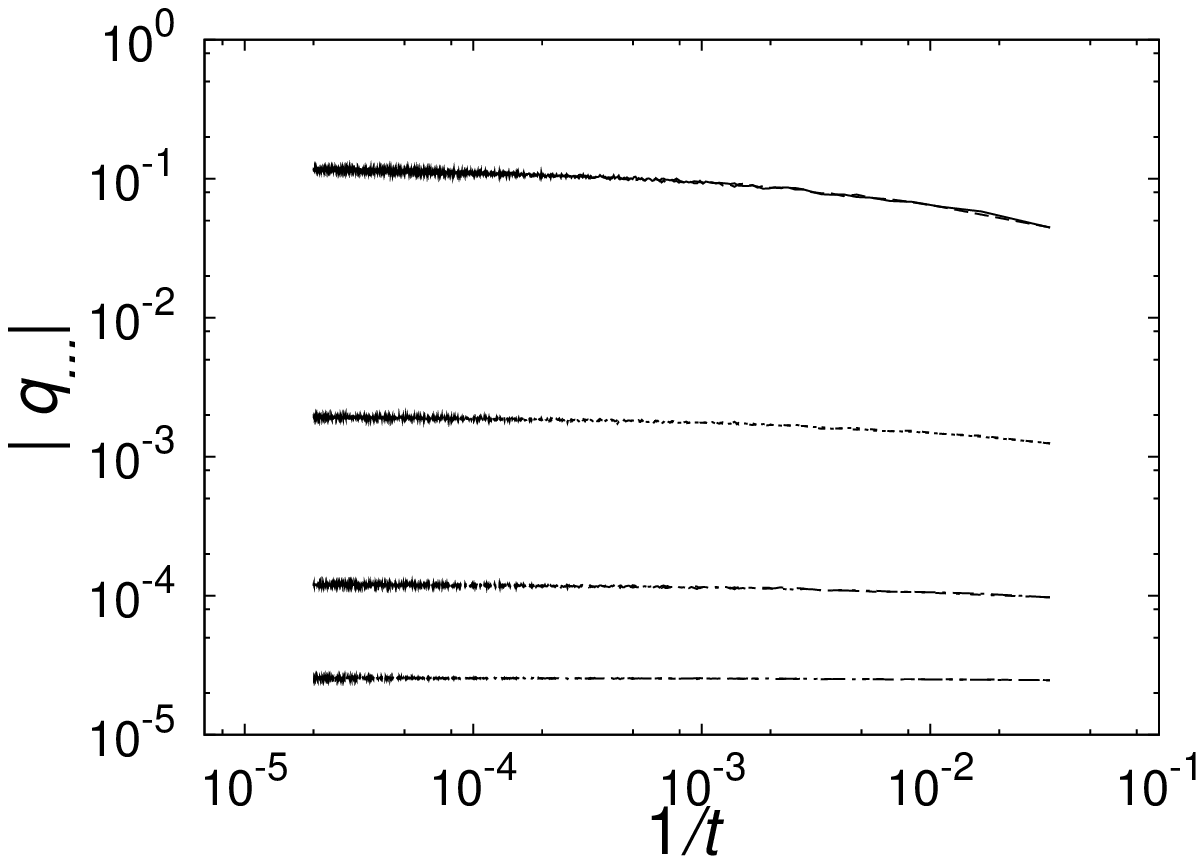} & \includegraphics[angle=0,width=0.45\columnwidth]{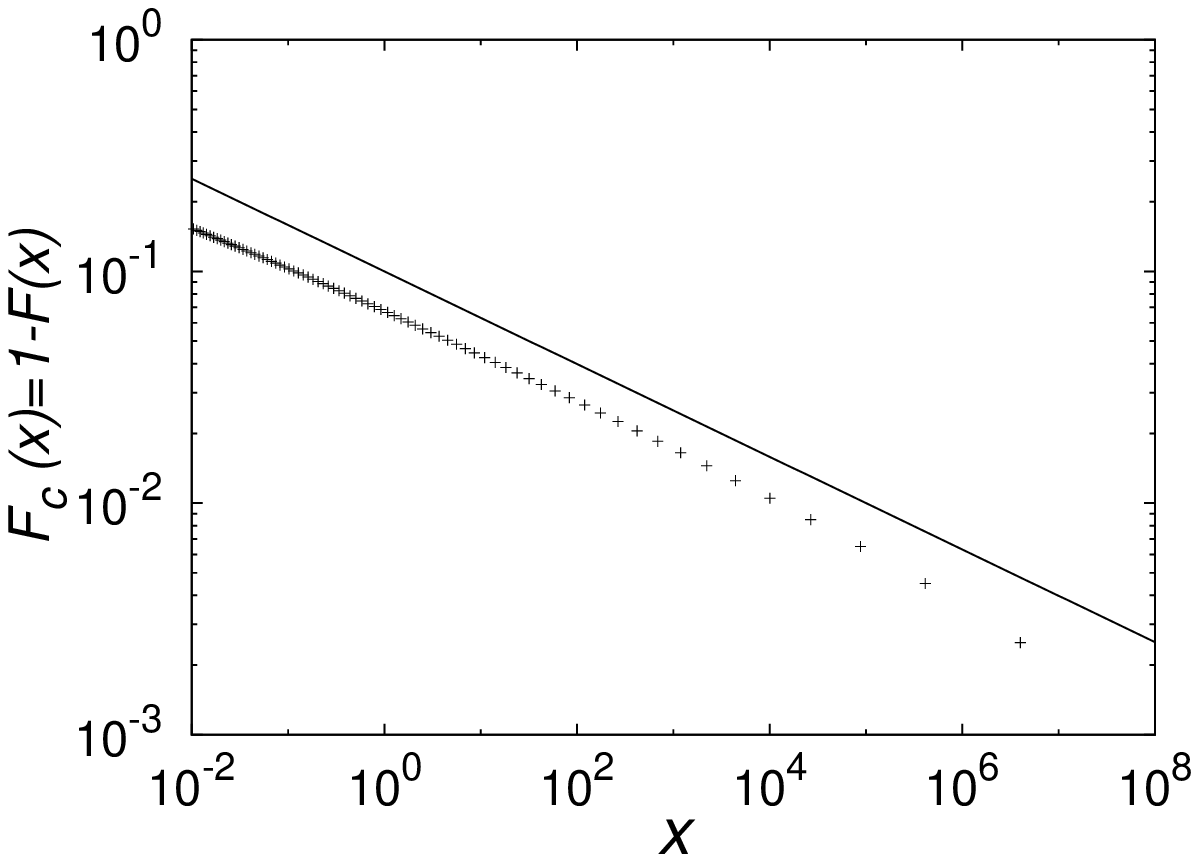}\\
\end{tabular}
\caption{Quantile lines $q_{y}$ and $|q_{1-y}|$
($y=\{0.9,0.8,0.7,0.6\}$, from top to bottom) as a function of $1/t$
(left column) and complementary cumulative distributions at the end
of simulation (right column) for $V(x)=|x|^{1.5}$ with the value of
the stability index $\alpha=0.7$ and the scale parameter $\sigma$:
$\sigma=1$ (top panel), $\sigma=0.1$ (middle panel) and
$\sigma=0.01$ (bottom panel). In the right column, solid lines
present $x^{-(\alpha-0.5)}$ decay predicted by
Eq.~(\ref{eq:hdfdecay}).} \label{fig:x15sigma}
\end{center}
\end{figure}

For the quartic potential $V(x)=x^4/4$ the expression for $\hat{U}(k)$ reads $\hat{U}(k)=k\frac{\partial^3}{\partial k^3}$. The stationary density fulfills
\begin{equation}
\frac{\partial^3\hat{P}(k)}{\partial k^3}=\sigma^\alpha\mathrm{sign}k|k|^{\alpha-1}\hat{P}(k).
\label{eq:quartic}
\end{equation}
For $\alpha=1$, the solution of Eq.~(\ref{eq:quartic}) reads \cite{chechkin2002,chechkin2003}
\begin{equation}
\hat{P}_{\alpha=1}(k)=\frac{2}{\sqrt{3}}\exp\left[-\frac{\sigma^{1/3}|k|}{2}\right]\cos\left[\frac{\sqrt{3}\sigma^{1/3}|k|}{2} -\frac{\pi}{6}\right].
\end{equation}
The formula for the corresponding stationary density $P(x)$ \cite{chechkin2002,chechkin2003,chechkin2004,chechkin2006} in the real space is
\begin{equation}
P_{\alpha=1}(x)=\frac{1}{\pi w\left[ 1-(x/ w)^2+(x/ w)^4 \right]},
\label{eq:cauchystationary}
\end{equation}
where $ w =\sigma^{1/3}$.
The stationary solution (\ref{eq:cauchystationary}) of Eq.~(\ref{eq:quartic}), as well as the steady state~(\ref{eq:parabolic}) with $\alpha<2$, is no longer of the Boltz\-mann-Gibbs type. This is typical for system driven by L\'evy white noises with the stability index $\alpha<2$ \cite{eliazar2003}.
The stationary state, see Eq.~(\ref{eq:cauchystationary}), has the asymptotic power-law dependence, i.e. $P(x) \propto |x|^{-4}$. Furthermore, the stationary solution~(\ref{eq:cauchystationary}) is bimodal, with modal values located at $x=\pm {\sigma^{1/3}}/{\sqrt{2}}$ \cite{chechkin2002,chechkin2003,chechkin2004}.


\subsection{Subharmonic potentials\label{sec:subharmonic}}

%
%
\begin{figure}
\begin{center}
\begin{tabular}{p{0.45\columnwidth}p{0.45\columnwidth}}
\includegraphics[angle=0,width=0.45\columnwidth]{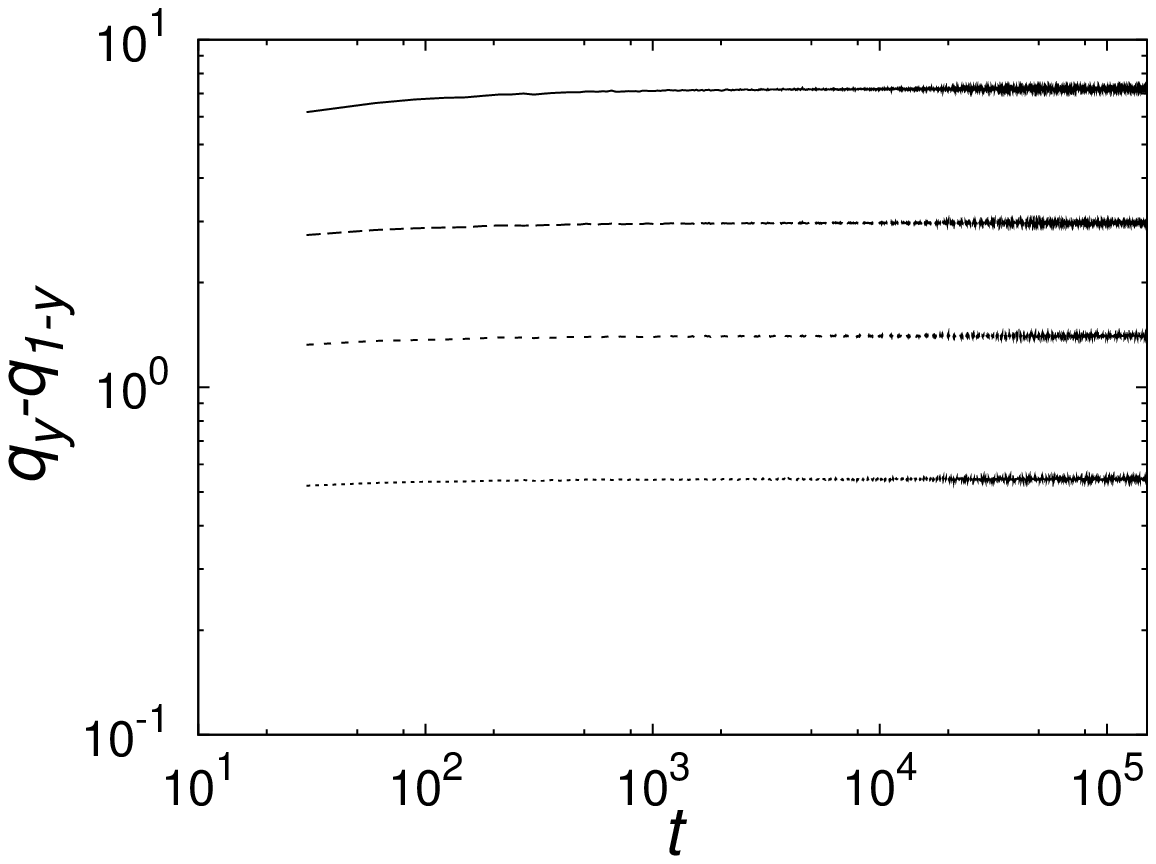} & \includegraphics[angle=0,width=0.45\columnwidth]{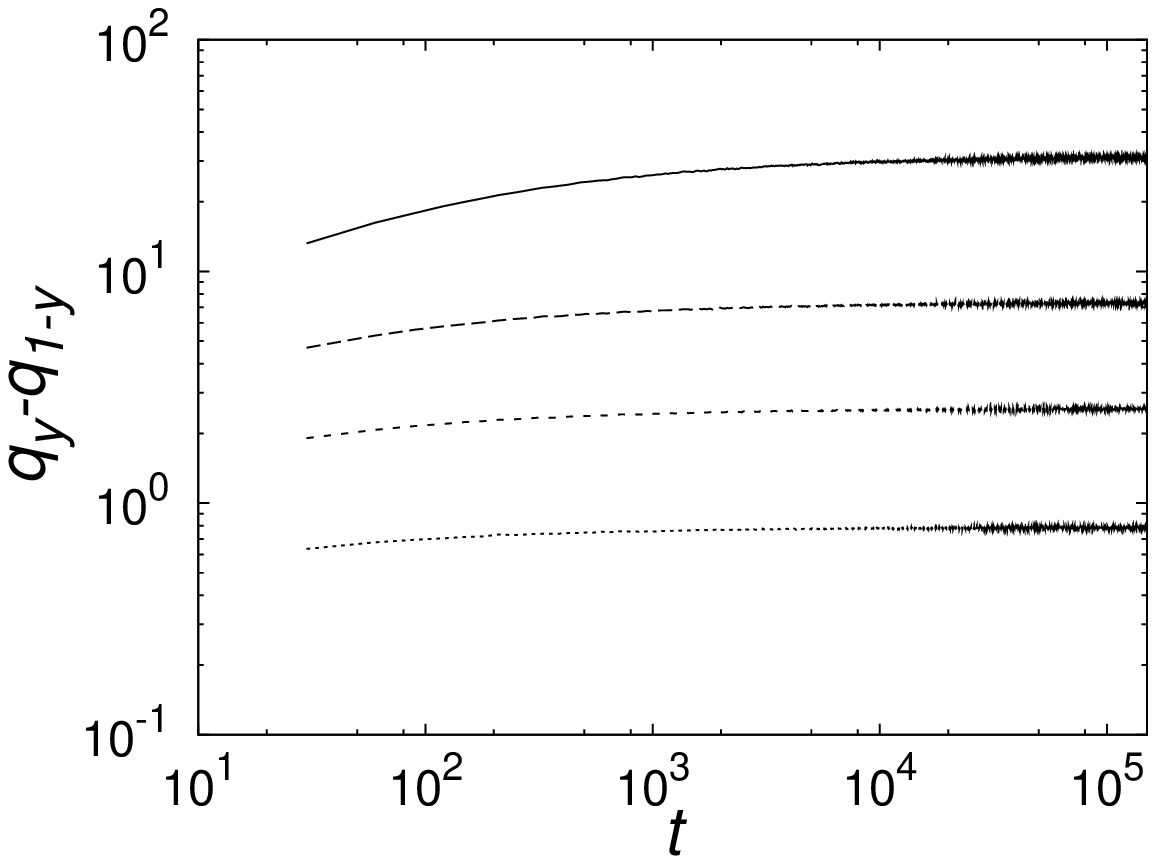}\\
\includegraphics[angle=0,width=0.45\columnwidth]{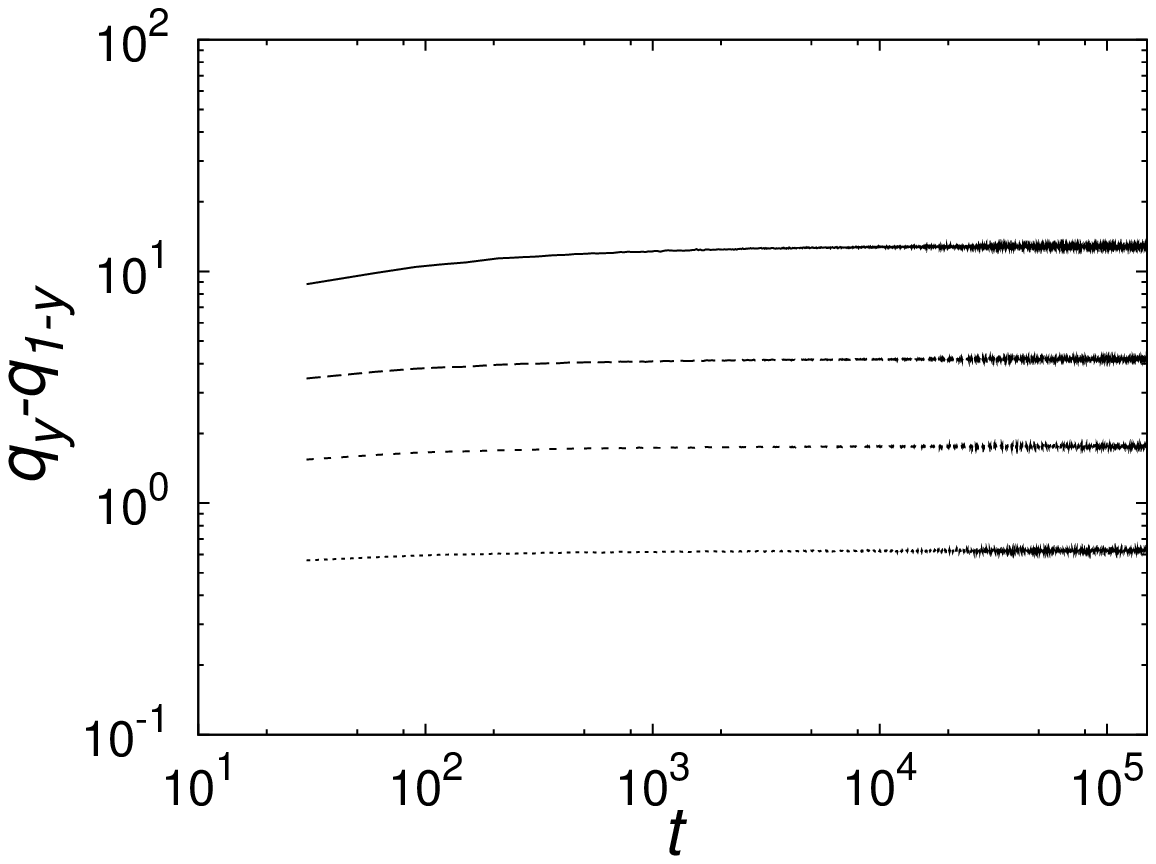} & \includegraphics[angle=0,width=0.45\columnwidth]{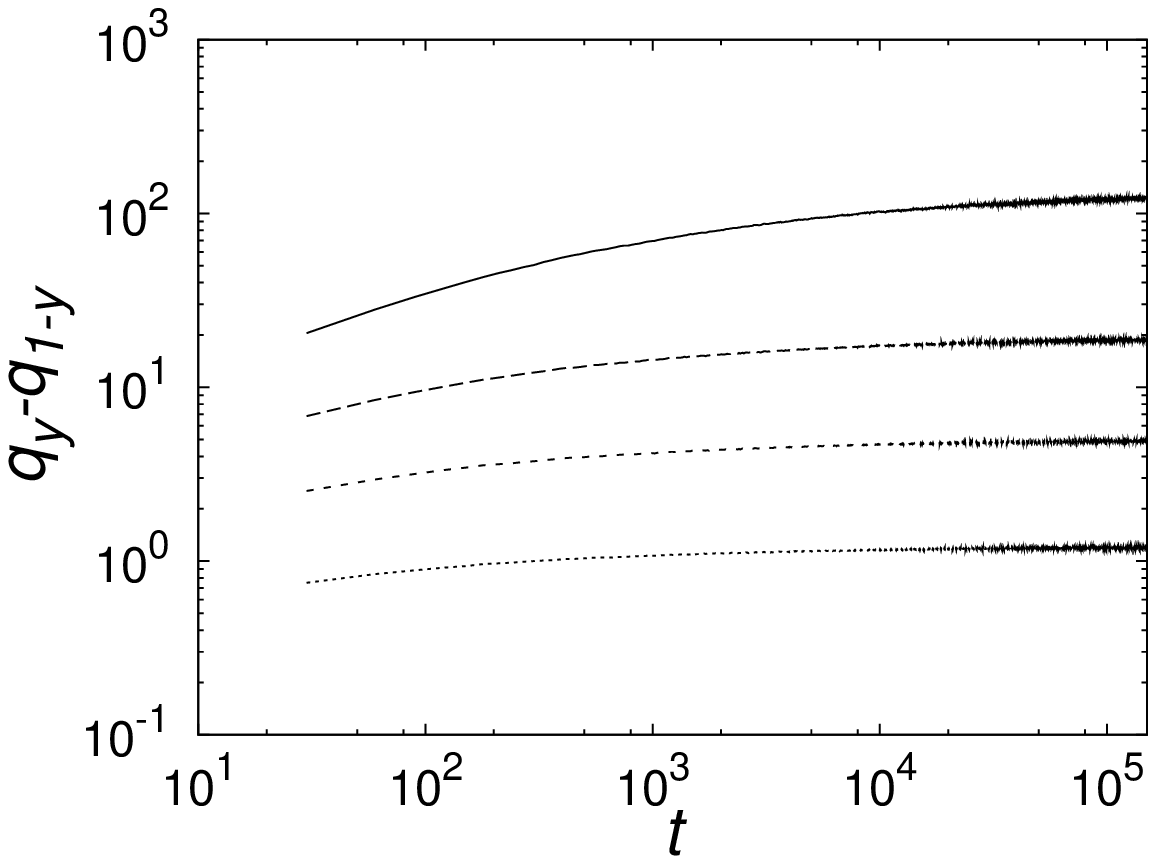}\\
\end{tabular}
\caption{Interquantile distance ($q_{0.9}-q_{0.1}$, $q_{0.8}-q_{0.2}$, $q_{0.7}-q_{0.3}$ and $q_{0.6}-q_{0.4}$, from top to bottom) as a function of time $t$  for $V(x)=|x|$ with different values of the stability index $\alpha=\{1.7,1.6\}$ (left column, from top to bottom) and $\alpha=\{1.5,1.4\}$ (right column, from top to bottom). The scale parameter $\sigma$ is set to $\sigma=1$.}
\label{fig:x1intqnt}
\end{center}
\end{figure}

\subsubsection{Analytical results.}

Subharmonic potentials ($V(x)=|x|^c$ with $0<c<2$, see
Fig.~\ref{fig:potential}) interpolate between the force free case
($c=0$) and the harmonic potential ($c=2$).  As it was indicated in
the previous subsection for the harmonic potential, the stationary
state is an $\alpha$-stable L\'evy type distribution, see
Eqs.~(\ref{eq:charakt}) and~(\ref{eq:parabolic}). In the force free
case the stationary solutions do not exist. The time dependent
solutions are just L\'evy stable distributions with the growing scale
parameter $\sigma$, i.e. $\sigma(t)=\sigma t^{1/\alpha}$.  This
observation suggests that for intermediate values of the exponent $c$
($0<c<2$) there should be a transition between the situation when
the stationary state exists and the situation when the stationary state is
absent.

It is possible to obtain a necessary condition for the
existence of the stationary state using qualitative arguments  based on the FFPE (\ref{eq:ffpe}).
If a stationary state for Eq.~(\ref{eq:ffpe}) exists, it satisfies
\begin{equation}
\left[ \frac{d}{d x} V'(x) + \sigma^\alpha \frac{d^\alpha}{d |x|^\alpha}\right]P(x)=0.
\label{eq:stationryffpe}
\end{equation}
Assuming that tails of the distribution are given by
a power-law, i.e. $P(x) \propto |x|^{-\nu}$ as $|x| \to \infty$, the first term in Eq.~(\ref{eq:stationryffpe}) behaves as
\begin{equation}
\frac{d}{d x} \left[c |x|^{c-1} P(x)\right] \simeq x^{c-2-\nu}.
\label{eq:deterministic}
\end{equation}
The fractional derivative, see Eq.~(\ref{eq:stationryffpe}), is a
nonlocal operator. The asymptotics of the fractional derivative of a
probability density has a universal behavior independent of the
particular form of this density. Indeed, its behavior for $x \to
\infty$ is governed by the behavior of its Fourier-transform
$-|k|^\alpha \hat{P}(k)$ for $k \to 0$. Due to the normalization
condition $\hat{P}(k=0)=1$. The inverse Fourier transform results in
\begin{equation}
\frac{d^\alpha}{d |x|^\alpha} P(x) \simeq x^{-1-\alpha}.
\label{eq:invfourier}
\end{equation}
In order to satisfy Eq.~(\ref{eq:stationryffpe}), both terms (\ref{eq:deterministic}) and (\ref{eq:invfourier}) have to represent the same $x$-dependence. Consequently, we get $c-2-\nu = -1 - \alpha$
and
\begin{equation}
\nu = c +\alpha  - 1.
\label{eq:decay}
\end{equation}
Therefore a stationary state is asymptotically characterized by a power-law
\begin{equation}
 P(x) \propto |x|^{-(c+\alpha-1)},
\end{equation}
i.e. by the same expression as for the superharmonic potentials \cite{chechkin2004}.
The probability density function $P(x)$ has to be integrable, which corresponds to $\nu > 1$.
Thus, the necessary condition for the existence of the steady state is
\begin{equation}
c > 2-\alpha
\label{eq:existencecondition}
\end{equation}
or
\begin{equation}
\alpha > 2-c.
\end{equation}

For the parabolic potential ($c=2$), the exponent $\nu$ reproduces the decay of the
L\'evy distribution ($\nu = 1+\alpha$), see Eq.~(\ref{eq:parabolic}).
For the quartic potential ($c=4$) with
$\alpha=1$ we get $\nu = 4$, as follows from Eq.~(\ref{eq:cauchystationary}).
The complementary cumulative density function ($F_c(x)=1-F(x)=1-\int_{-\infty}^xP(x')dx'$)
is thus asymptotically characterized by a power-law decay
\begin{equation}
 F_c(x) \propto x^{-(c+\alpha-2)}.
\label{eq:hdfdecay}
\end{equation}

Having the condition (\ref{eq:existencecondition}), it is also possible to determine the dependence of the width of the stationary distribution $w(\sigma)$ on the noise strength $\sigma$.
Assuming that $P(x) = w^{-1} f(x/w)$ can be expressed through a universal function
$f(\xi)$ of a new, dimensionless, length variable $\xi = x/w$ (this assumption
is reasonable due to the scale-free nature of the power-law potential), we
change the length variable to $\xi=x/w$.
For $V(x) = |x|^c$ (i.e.  $V'(x) \propto c |x|^{c-1}$, $x\neq 0$), we get
\begin{equation}
w^{c-1} w^{-2} \frac{d}{d \xi} \left[c |\xi|^{c-1} f(\xi) \right] + \sigma^\alpha w^{-1-\alpha} \frac{d^\alpha}{d |\xi|^\alpha} f(\xi) =0.
\label{eq:rescalled}
\end{equation}
The universal rescaled equation for $f(\xi)$
\begin{equation}
 \frac{d}{d \xi} \left[c |\xi|^{c-1} f(\xi) \right] +
 \frac{d^\alpha}{d |\xi|^\alpha} f(\xi) =0
\label{eq:dimensionless}
\end{equation}
can only be obtained if the width of distribution  $w(\sigma)$ is the solution of the following equation
\begin{equation}
w^{c-3} = \sigma^\alpha  w^{-1-\alpha}.
\label{eq:prefactors}
\end{equation}
Condition (\ref{eq:prefactors}) assures that prefactors of both terms in Eq.~(\ref{eq:rescalled}) scale similarly when $\sigma$ is changed. Thus, from Eq.~(\ref{eq:prefactors}) one obtains
\begin{equation}
w(\sigma) \propto \sigma^{\alpha/(c+\alpha-2)}.
\label{eq:width}
\end{equation}

In the absence of the noise ($\sigma \to 0$) the solution $P(x)$ corresponds to the particle's
position in the minimum of the potential, $P(x)=\delta(x)$, i.e. $w(\sigma)=0$.
For $c+\alpha <2$ the exponent in Eq.~(\ref{eq:width}) is negative, and
the distribution's width $w(\sigma)$ diverges for $\sigma \to 0$ which corresponds to a non-physical situation. Consequently,
no stationary state is possible. On the contrary, for $c+\alpha >2$ the exponent in Eq.~(\ref{eq:width}) is positive and the distribution width is the increasing function of the scale parameter $\sigma$. Therefore, the necessary condition of the existence of the steady state, see Eq.~(\ref{eq:existencecondition}), is confirmed by the condition (\ref{eq:width}).

Note that the cases of parabolic and quartic potentials confirm Eq.~(\ref{eq:width}).
For the parabolic potential ($c=2$) we get $w(\sigma) \propto \sigma$, see Eq.~(\ref{eq:parabolic}).
In the case of quartic potential ($c=4$) with $\alpha = 1$ we get $w(\sigma) \propto \sigma^{1/3}$, see Eq.~(\ref{eq:cauchystationary}).
Moreover, it is also possible to discuss how does the relaxation time to the equilibrium
depends on the scale parameter $\sigma$. To do this, we turn to Eq.~(\ref{eq:ffpe}) and rescale both the length ($\xi=x/w$) and the time variables ($\tau = t/\tau_0(\sigma)$), so
that the rescaled solution to Eq.~(\ref{eq:ffpe}) is a universal function.
The rescaling of $t$ corresponds to
\begin{equation}
\tau_0(\sigma) \propto \left[ w(\sigma) \right]^{3-c} = \sigma^{\alpha(3-c)/(c+\alpha-2)}.
\end{equation}
It indicates that the dependence on the scale parameter $\sigma$ for
the stability index $\alpha$ close to $2-c$ gets extremely sharp. In
other words, when $\alpha$ approaches $2 - c$ (from above) the
relaxation time grows and diverges at $\alpha = 2 - c$.

The condition (\ref{eq:existencecondition}) for the existence of the
steady state can be corroborated by the following qualitative argument
based on the decomposition of the noise into a ``background'' noise
with finite variance, and ``bursts'' of arbitrary amplitude
\cite{imkeller2006,imkeller2006b}.

Let us consider the discretized Langevin scheme with the time step
of integration $\Delta t$ and the characteristic amplitude of the
L\'evy jumps $\epsilon = \sigma (\Delta t)^{1/\alpha}$.
Additionally, let us introduce a cutoff level $\Lambda$, so that
only the jumps with the amplitude $a > \Lambda$ are considered as
bursts. The overall situation is then considered as pertinent to a
perturbation of the equilibrium distribution of particles in the
potential $V(x)$ under the action of the background noise (i.e. the
Boltzmann distribution in the potential $V(x)$ at the temperature $kT
\simeq \Lambda$) by these distinct bursts. The width of this
distribution $W$ depends on the cutoff level $\Lambda$ and on the
type of the potential. Moreover, we assume that only very large
bursts, with $a \gg \Lambda$, can be responsible for the absence of
a stationary distribution. Therefore, the finite width of the
background distribution can be neglected when considering the action
of such bursts. The distribution of the burst amplitudes is then
given by the asymptotic behavior of tails of the symmetric L\'evy
distribution
\begin{equation}
p(a) \simeq \frac{\epsilon^\alpha}{x^{1+\alpha}} = \frac{\sigma^\alpha \Delta t}{x^{1+\alpha}} .
\end{equation}

A burst of amplitude $a$ brings the particle to a position $x \approx a$, and the
typical time necessary
to return to the interval of the characteristic width $W$ is given by the solution of the ordinary
equation of motion
\begin{equation}
\dot{x}(t)=-V'(x).
\end{equation}
For $c\neq 2$, the return time $T(a)$ into the equilibrium domain, i.e. $|x|<W$, from the initial position $x=a$ is given by
\begin{equation}
T(a)=\frac{1}{c(2-c)}\left[a^{2-c}-W^{2-c}\right].
\label{ret1}
\end{equation}
For subharmonic potentials, $0<c<2$, the return time $T(a)$ is
dominated by the initial position $a$. The leading contribution to
$T(a)$ is provided by the first term of Eq.~(\ref{ret1}),
i.e. $a^{2-c}/[c(2-c)]$. Therefore, instead of the return time to the
equilibrium domain it is possible to consider the return time to the
origin, which is finite and slightly larger than the return time to
the characteristic domain.

It is reasonable to assume, that the steady state does not exist, if
the time $T(a)$ to return from the point $a$ is larger than the
typical time between two bursts of the amplitude $a$ or larger, and
might exist in the opposite situation. The probability to have a
burst of the amplitude $a$ or larger is
\begin{equation}
P(a) = 2 \int_a^\infty p(a') da' \simeq \left(\frac{\epsilon}{a} \right)^{\alpha},
\end{equation}
so that the typical time between two such events is
\begin{equation}
T \simeq \Delta t \left(\frac{a}{\epsilon} \right)^{\alpha}.
\label{eq:returntime}
\end{equation}
Comparing $T$ with $T(a)$ we conclude that
the inequality $T(a) < T$ applies and the stationary states are
possible for
\begin{equation}
c > 2-\alpha.
\end{equation}

For superharmonic potentials, $c>2$, the situation is inverse to the
situation for subharmonic potentials. The larger the initial
displacement $a$ is, the faster is the return to the $W$-domain. For
the increasing initial displacement $a$, the typical return time
essentially tends to a constant value $W^{2-c}/[c(c-2)]$, while the
time between two subsequent bursts grows with their amplitude
$a$. Comparing $T$ with $T(a)$, we conclude that stationary states for
superharmonic potentials exist for every value of the stability index
$\alpha$. The harmonic potential, $c=2$, is the limiting case for
which the return time shows not a power-law but a logarithmic
dependence. In such a case, the very same arguments proofing existence
of stationary states hold.

\subsubsection{Numerical results.}

Numerical results were obtained by Monte Carlo simulations of
Eq.~(\ref{eq:langevin}) with the time step of integration $\Delta
t=10^{-2}$ and averaged over $N=10^6$ realizations
\cite{janicki1994,janicki1996,janicki2001,dybiec2004b,dybiec2006}.
Initially, at time $t=0$ a test particle was located at the origin,
i.e. $x(0)=0$.  Due to the symmetry of the noise and of the potential,
stationary states are symmetric with respect to the origin; the median
and modal values of stationary densities are located at the origin.

Our test of stationarity is based on the quantiles of
distributions. Quantiles of stationary distributions remain
constant. Consequently, plotting quantile values as functions of time (quantile
lines) gives us the key to analyze stationarity. More precisely, if the
quantile lines are parallel to the abscissa, it means that the stationary
state has been reached.

In order to verify performance of the test based on quantile lines we
have constructed quantile lines for L\'evy flights in the parabolic
potential. In such a case the stationary state exists and is given by the
$\alpha$-stable density, see Eq.~(\ref{eq:parabolic}). Therefore,
quantile lines should be parallel to the abscissa. Analogously,
interquantile distance should not change over time. This behavior is
indeed observed, e.g. in the bottom panel of
Fig.~\ref{fig:potentials}.

Furthermore, Fig.~\ref{fig:potentials} compares interquantile
distance (left column) and complementary cumulative distributions
$F_c(x,t)=1-F(x,t)=1-\int_{-\infty}^xP(x',t)dx'$ (right column) for
the stability index $\alpha=0.9$ and various potentials: $V(x)=0$
(top panel), $V(x)=|x|$ (middle panel) and $V(x)=x^2/2$ (bottom
panel). From these cases only the $V(x)=x^2/2$ case corresponds to
the situation when stationary state exists. Top panel of
Fig.~\ref{fig:potentials} corresponds to the force free case in
which no stationary density exists. In such a case, the
interquantile distance grows like $t^{1/\alpha}$. The middle panel
of Fig.~\ref{fig:potentials} corresponds to the intermediate case,
i.e. $V(x)=|x|$. Numerical simulations indicate that for $V(x)=|x|$
and $\alpha=0.9$ no stationary state exists as well, what is
coherent with analytical arguments given above. Note that in all
three cases the decay of the probability densities follows the one
of the noise, albeit on different reasons. For free motion, this
decay follows the one of the noise due to the stability of the
process. For the case $V(x)=|x|$, where according to analytical
result the stationary state is absent, numerical simulations suggest
that the tail of the distribution function is still dominated by the
properties of the noise, and follows  $F_c(x) \propto x^{-\alpha}$
pattern. Finally, for the parabolic potential it is essentially
given by $F_c(x) \propto x^{-(c+\alpha-2)}$, i.e. again by $F_c(x)
\propto x^{-\alpha}$ for $c=2$.


%
%
\begin{figure}
\begin{center}
\begin{tabular}{p{0.45\columnwidth}p{0.45\columnwidth}}
\includegraphics[angle=0,width=0.45\columnwidth]{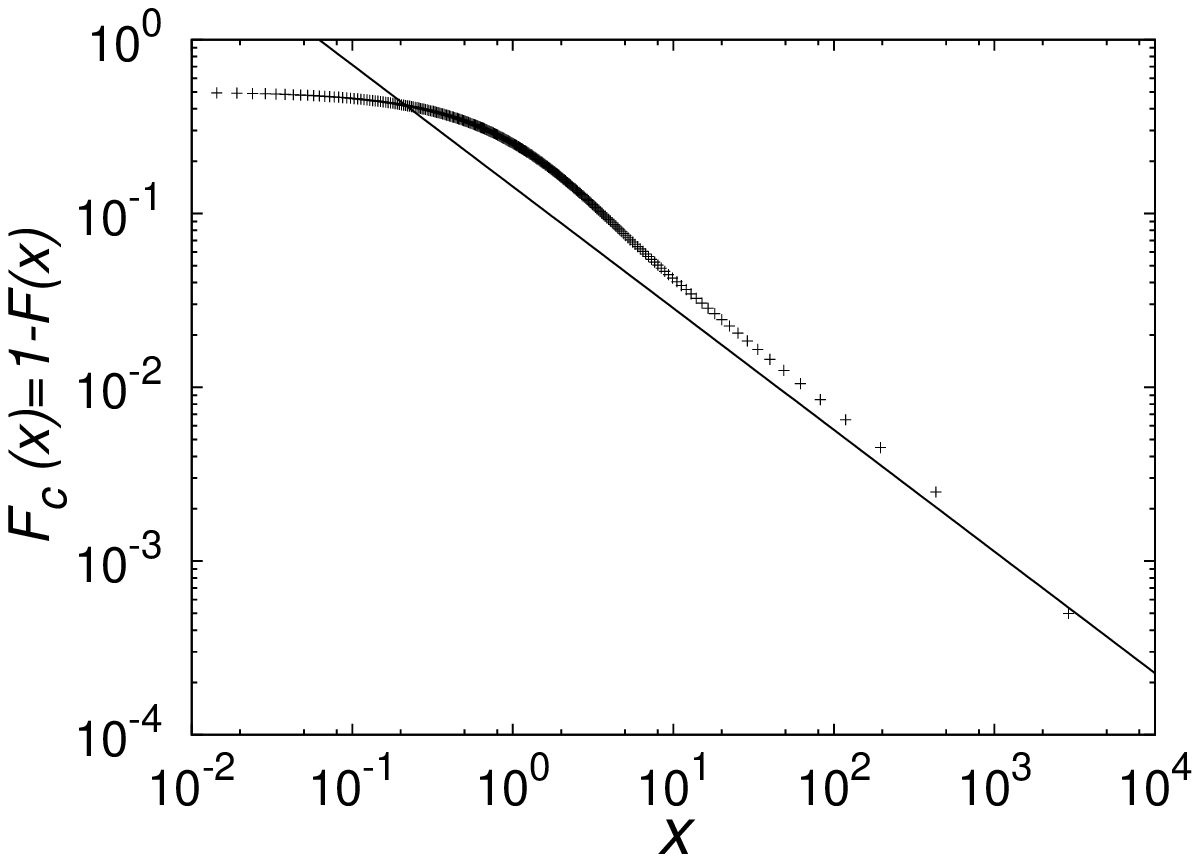} & \includegraphics[angle=0,width=0.45\columnwidth]{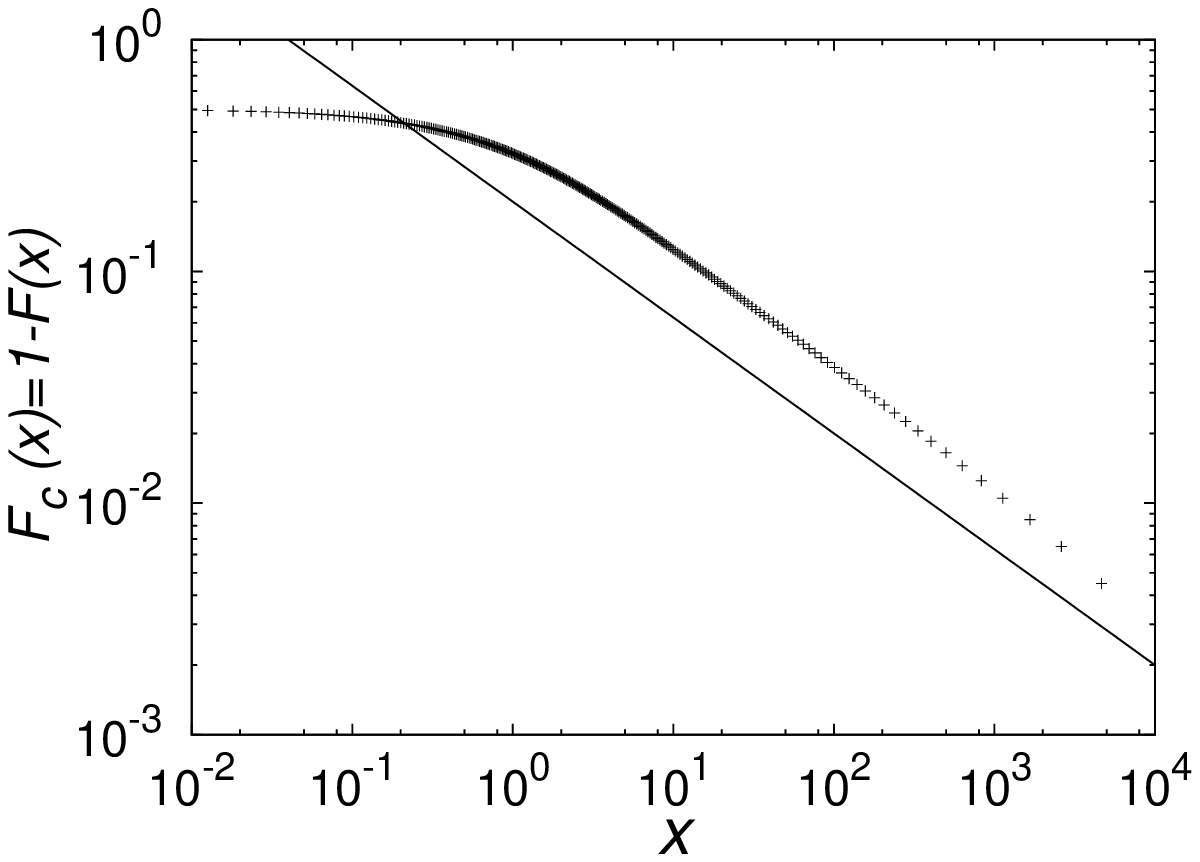}\\
\includegraphics[angle=0,width=0.45\columnwidth]{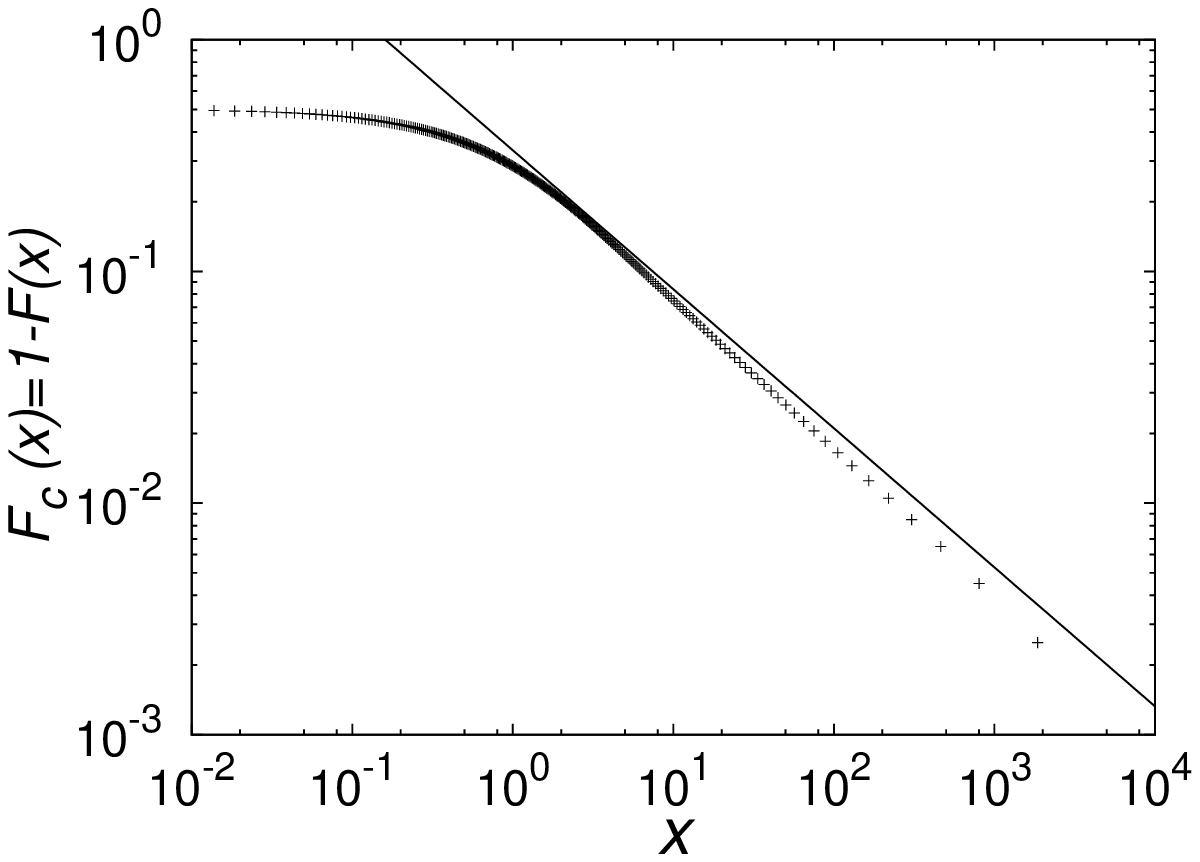} & \includegraphics[angle=0,width=0.45\columnwidth]{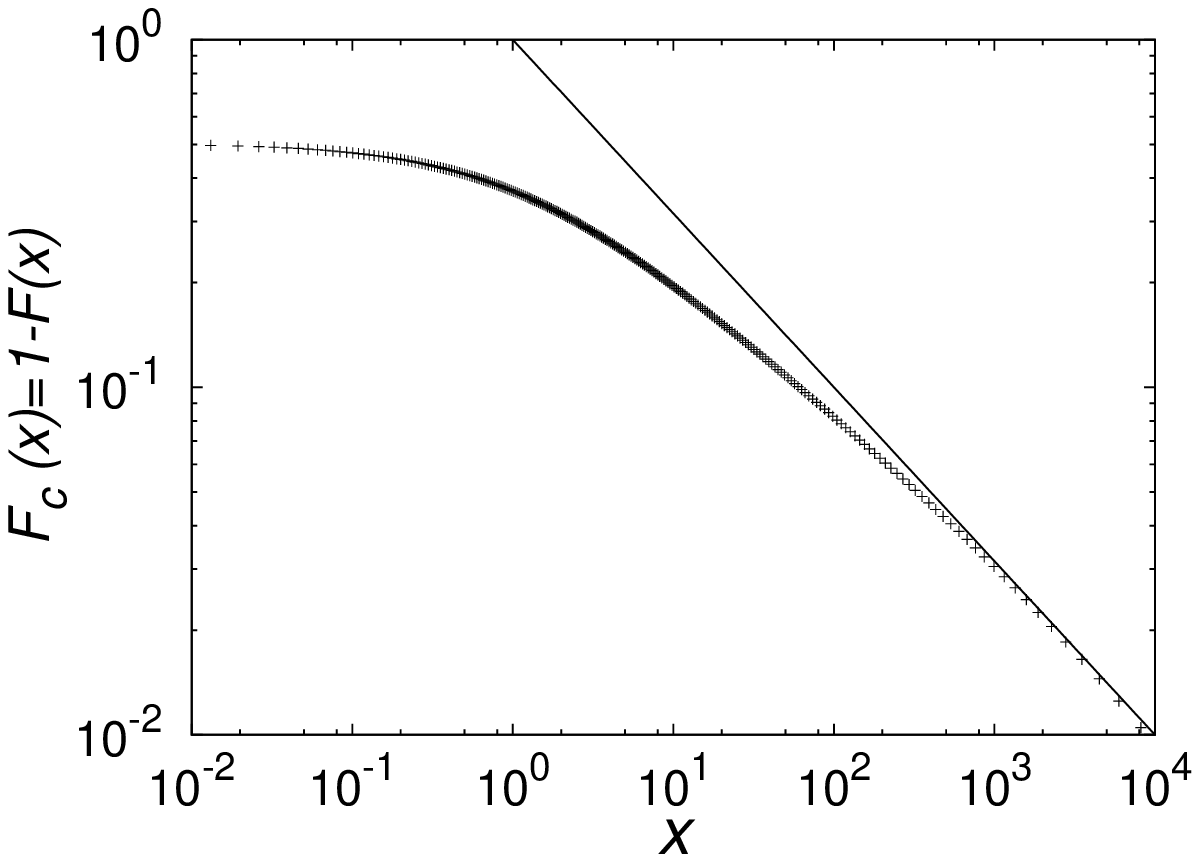}\\
\end{tabular}
\caption{Complementary cumulative distributions at the end of
simulation for $V(x)=|x|$ with different values of the stability
index $\alpha=\{1.7,1.6\}$ (left column, from top to bottom) and
$\alpha=\{1.5,1.4\}$ (right column, from top to bottom). The scale
parameter $\sigma$ is set to $\sigma=1$. Solid lines present
$x^{-(\alpha-1)}$ decay predicted by Eq.~(\ref{eq:hdfdecay}).}
\label{fig:x1cdf}
\end{center}
\end{figure}

Consecutive figures present results for subharmonic potentials.
Figs.~\ref{fig:x15intqnt}--\ref{fig:x15sigma} present results for
$V(x)=|x|^{1.5}$. Figs.~\ref{fig:x15intqnt}--\ref{fig:x15cdf}
present: interquantile distance as a function of time $t$
(Fig.~\ref{fig:x15intqnt}) and complementary cumulative
distributions ($F_c(x)=1-F(x)$) at the end of simulation
(Fig.~\ref{fig:x15cdf}). Various panels present result for various
values of the stability index $\alpha$: $\alpha=\{1.5,1.1\}$ (left
column, from top to bottom) and $\alpha=\{0.9,0.8\}$ (right column,
from top to bottom). Finally, Fig.~\ref{fig:x15sigma} presents
quantile lines (left column) and complementary cumulative
distributions (right column) for $\alpha=0.7$ with various scale
parameters $\sigma$: $\sigma=1$ (top panel), $\sigma=0.1$ (middle
panel) and $\sigma=0.01$ (bottom panel).

\begin{figure}
\begin{center}
\begin{tabular}{p{0.45\columnwidth}p{0.45\columnwidth}}
\includegraphics[angle=0,width=0.45\columnwidth]{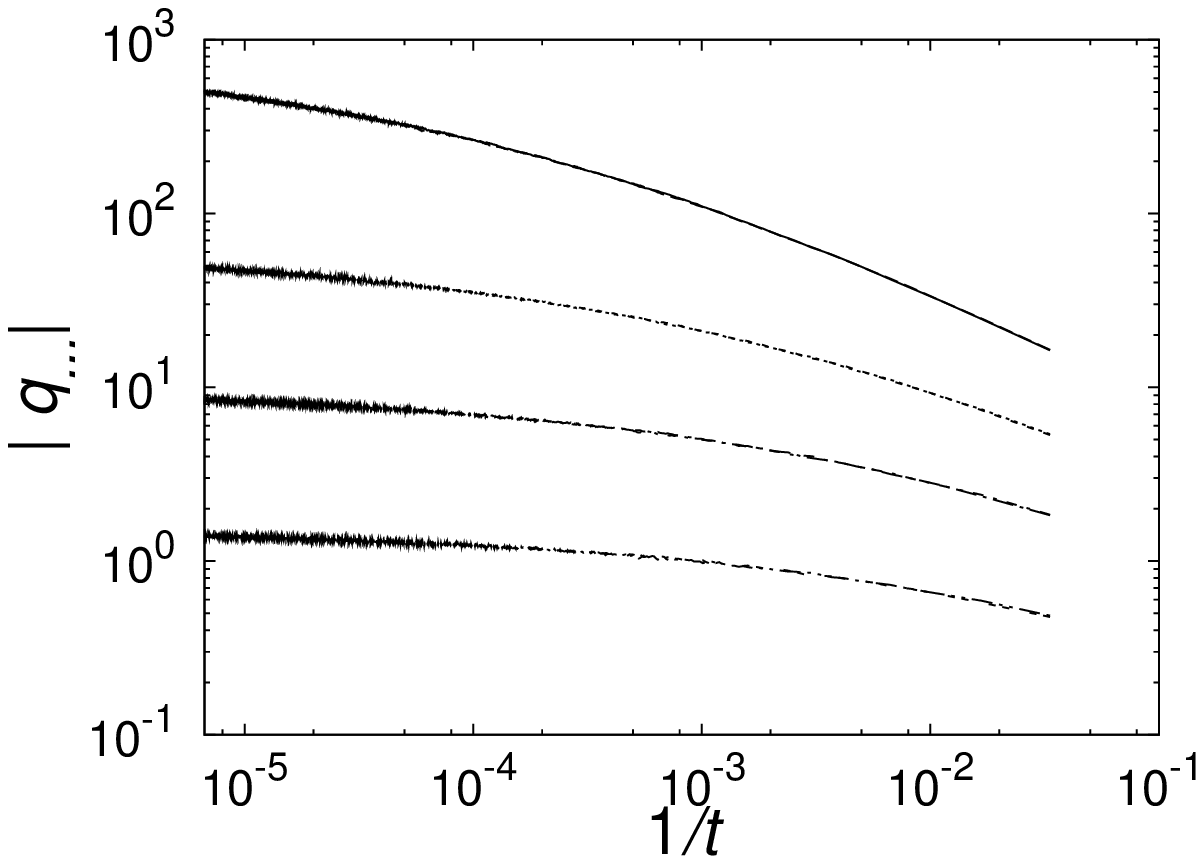} & \includegraphics[angle=0,width=0.45\columnwidth]{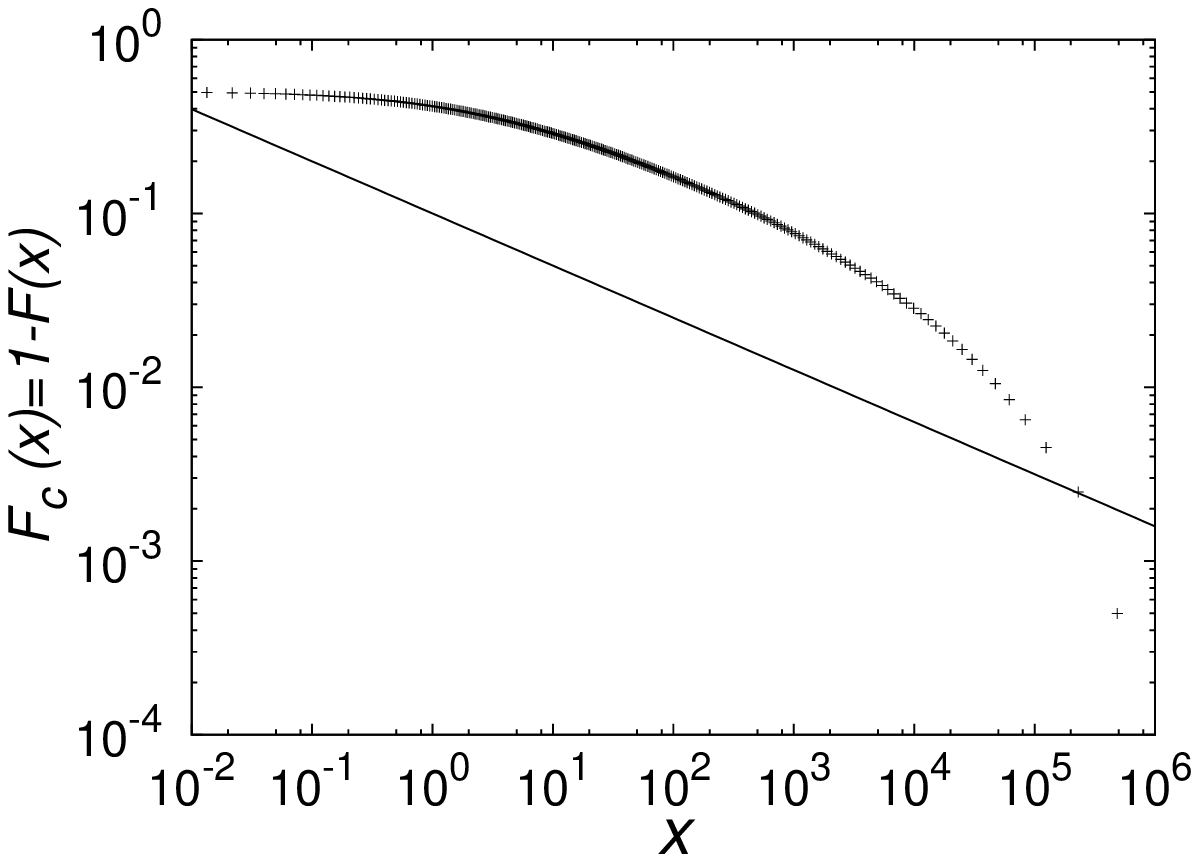}\\
\includegraphics[angle=0,width=0.45\columnwidth]{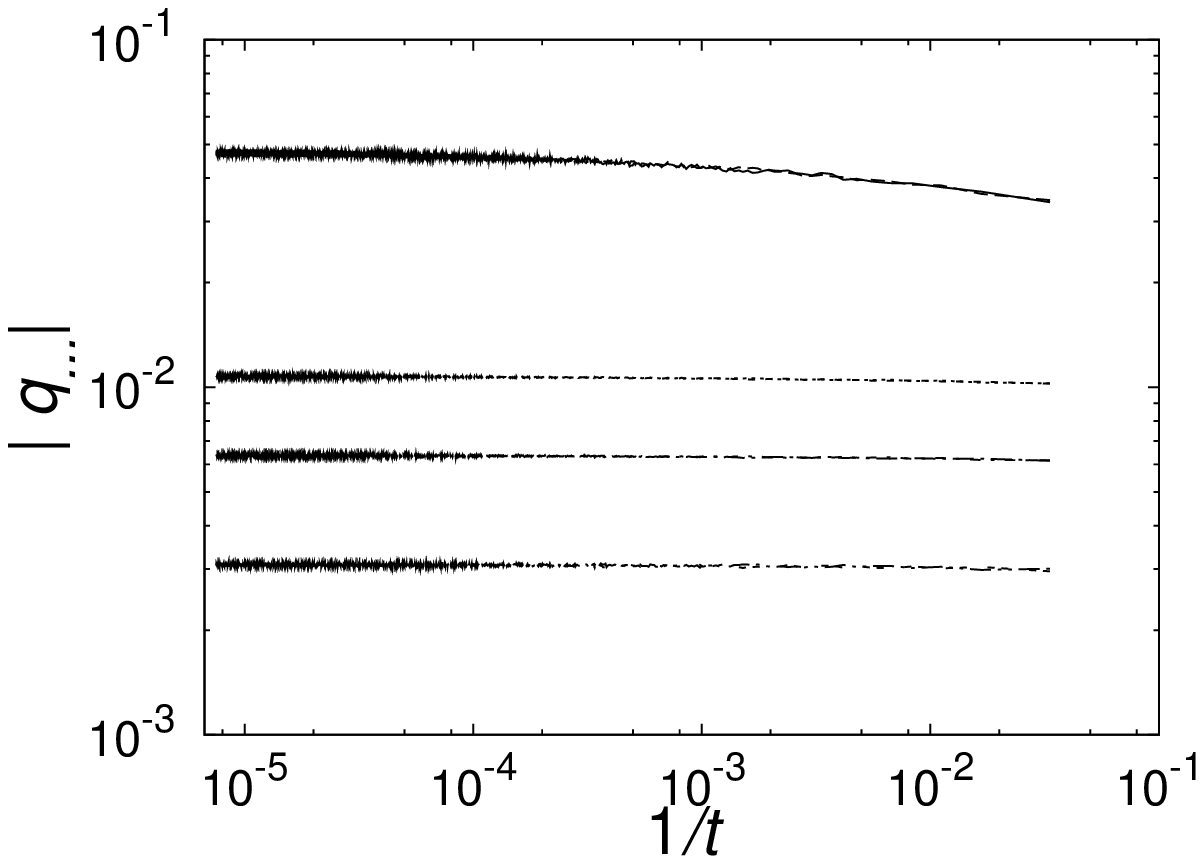} & \includegraphics[angle=0,width=0.45\columnwidth]{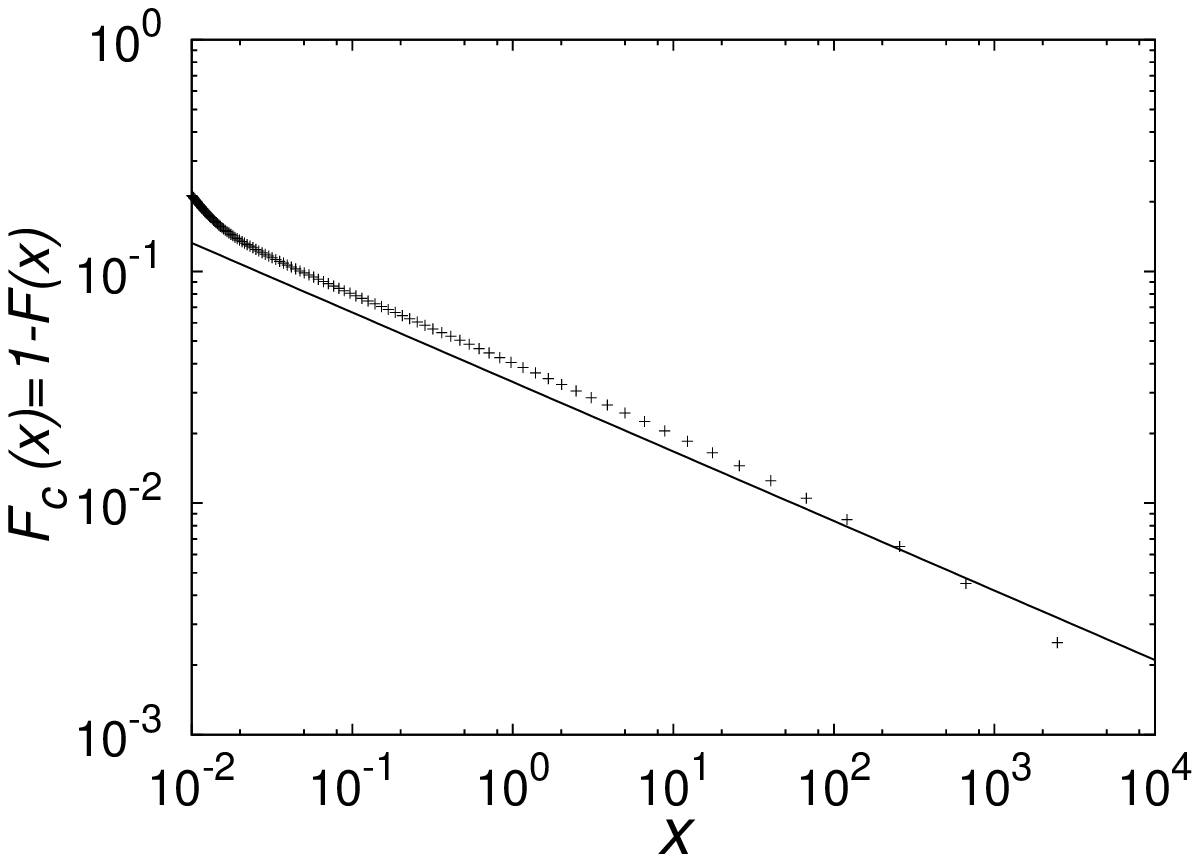}\\
\end{tabular}
\caption{Quantile lines $q_{y}$ and $q_{1-y}$
($y=\{0.9,0.8,0.7,0.6\}$, from top to bottom) as a function of $1/t$
(left column) and complementary cumulative distributions at the end
of simulation (right column) for $V(x)=|x|$ with the value of the
stability index $\alpha=1.3$ and the scale parameter $\sigma$:
$\sigma=1$ (top panel) and $\sigma=0.1$ (bottom panel). In the right
column, solid lines present $x^{-(\alpha-1)}$ decay predicted by
Eq.~(\ref{eq:hdfdecay}).} \label{fig:x1sigma}
\end{center}
\end{figure}

Figures~\ref{fig:x1intqnt}--\ref{fig:x1sigma} present results for
$V(x)=|x|$. Figs.~\ref{fig:x1intqnt}--\ref{fig:x1cdf} present:
interquantile distance as a function of time $t$
(Fig.~\ref{fig:x1intqnt}) and complementary cumulative distributions
($F_c(x)=1-F(x)$) at the end of simulation (Fig.~\ref{fig:x1cdf}).
Various panels present result for various values of the stability
index $\alpha$: $\alpha=\{1.7,1.6\}$ (left column, from top to
bottom) and $\alpha=\{1.5,1.4\}$ (right column, from top to bottom).
Finally, Fig.~\ref{fig:x1sigma} presents quantile lines (left
column) and complementary cumulative distributions (right column)
for $\alpha=1.3$ with various scale parameters $\sigma$: $\sigma=1$
(top panel) and $\sigma=0.1$ (bottom panel).

The stability index $\alpha$ in
Figs.~\ref{fig:x15intqnt}--\ref{fig:x1sigma} is chosen in such a way that
Eq.~(\ref{eq:existencecondition}) holds true. Even in this case the
numerical verification whether stationary states exist or do not
exist, due to possible slow convergence to a stationary state, is
not trivial, see discussion of Figs.~\ref{fig:x15sigma} and
\ref{fig:x1sigma} below.

The probability densities for the system described by
Eqs.~(\ref{eq:langevin}) and (\ref{eq:ffpe}) are symmetric with
respect to the origin. Therefore the quantile lines for $q_y$ and
$|q_{1-y}|$ overlap and are not distinguishable in left panels of
Figs.~\ref{fig:x15sigma} and~\ref{fig:x1sigma}.

For $V(x)=|x|^{1.5}$, Fig.~\ref{fig:x15intqnt} indicates that for
sufficiently large values of the stability index $\alpha$ stationary
densities exist. Although all the values of stability indices shown
correspond to the domain where stationary states are predicted to
exist, the convergence to these states becomes slower when $\alpha$
approaches $2-c = 0.5$ for $V(x)=|x|^{1.5}$. Fig.~\ref{fig:x1intqnt}
shows the very similar behavior for $V(x)=|x|$. Here again the
relaxation time to the equilibrium gets larger when $\alpha$
approaches $2-c = 1$.

Figures~\ref{fig:x15cdf} and \ref{fig:x1cdf} present complementary
cumulative distributions in the situation when
Eq.~(\ref{eq:existencecondition}) is valid. Therefore, according to
Eq.~(\ref{eq:hdfdecay}), complementary cumulative distributions
should asymptotically decay as $x^{-(c+\alpha-2)}$. Indeed, in
Figs.~\ref{fig:x15cdf} and \ref{fig:x1cdf} the decay predicted by
Eq.~(\ref{eq:hdfdecay}) is observed.

Figures~\ref{fig:x15sigma} ($\alpha=0.7$ and $V(x)=|x|^{1.5}$) and
\ref{fig:x1sigma} ($\alpha=1.3$ and $V(x)=|x|$) present quantile
lines and complementary cumulative distributions for different
values of the scale parameter $\sigma$. The quantile lines are
plotted as functions of $1/t$ in order to elucidate convergence.
Thus, in Fig.~\ref{fig:x15sigma} the convergence of $q_{0.9}$ is not
achieved over the time of simulation for $\sigma = 1$, but is
approached for $\sigma = 0.01$. Correspondingly, the tail of the
complementary cumulative density differs from theoretical
predictions for $\sigma = 1$ and approaches its theoretical
asymptotics for $\sigma = 0.01$. The same is true for the data in
Fig. \ref{fig:x1sigma}, where the convergence is achieved already
for $\sigma = 0.1$. This indicates that the problem of the slow
convergence to the stationary state has to be taken seriously.

%
%

\section{Summary and conclusions\label{sec:summary}}

We considered the problem of existence of stationary states in
power-law subharmonic potentials $V(x)=|x|^c$, $c<2$, under the
action of L\'evy-stable noise characterized by the stability index
$\alpha$. The scaling analysis of the fractional Fokker-Planck
equation leads us to the conclusion that such states exist if
$\alpha>2-c$. This conclusion is corroborated by an alternative
argument based on the decomposition of the L\'evy noise. For subharmonic potentials, the probability
density functions of stationary states are characterized by the
asymptotic power-law decay $P(x) \propto
|x|^{-\nu}=|x|^{-(c+\alpha-1)}$ for $|x| \to \infty$, which is
slower than decay of the tails of the corresponding L\'evy
distributions. Monte Carlo simulations of the Langevin equation
confirm these analytical findings, at least for $\alpha$ not too
close to $2-c$. The convergence to the stationary state for $\alpha$
approaching $2-c$ is very slow, which poses considerable numerical
problems, since both long times and very small values of scaling
parameter $\sigma$ are required. However, for $\sigma$ small enough
the tendency to converge is still observable.

\begin{acknowledgments}
The authors acknowledge the financial support by DFG within SFB555.
AVC acknowledges financial support from European Commission via MC IIF, grant 219966 LeFrac.
Computer simulations have been performed at Institute of Physics, Jagellonian University and
Academic Computer Center, Cyfronet AGH.
\end{acknowledgments}


\end{document}